\documentstyle[aps,prl,epsf,floats,multicol,amssymb,preprint,tighten]{revtex}

\begin{document}

\newcommand{\fig}[2]{\epsfxsize=#1\epsfbox{#2}} \reversemarginpar 
\newcommand{\mnote}[1]{$^*$\marginpar{$^*$ {\footnotesize #1}}}

\bibliographystyle{prsty}

\title{Reaction Diffusion
Models in One Dimension with Disorder} 

\author{Pierre Le Doussal}
\address{CNRS-Laboratoire de Physique Th\'eorique de l'Ecole\\
Normale Sup\'erieure, 24 rue Lhomond, F-75231
Paris}
\author{C\'ecile Monthus}
\address{Laboratoire de Physique Th\'eorique et Mod\`eles Statistiques \\
DPT-IPN, B\^at. 100, 91400 Orsay, France}
\maketitle
\begin{abstract}
We study a large class of one dimensional reaction
diffusion models with quenched disorder using
a real space renormalization group method (RSRG) which yields
exact results at large time.
Particles (e.g. of several species) undergo diffusion
with random local bias (Sinai model) and may react upon meeting.
We obtain a detailed description of the asymptotic 
states (i.e attractive fixed points of the RSRG),
such as the large time decay of the density of
each specie, their associated universal amplitudes, 
and the spatial distribution of particles.
We also derive the spectrum of non trivial exponents which
characterize the convergence towards the asymptotic states.
For reactions which lead to several possible asymptotic
states separated by unstable fixed points, we analyze the
dynamical phase diagram and obtain the critical exponents 
characterizing the transitions. We also obtain a detailed
characterization of the persistence properties
for single particles as well as more complex patterns.
We compute the decay exponents for the probability of
no crossing of a given point by, respectively, the single particle 
trajectories ($\theta$) or the thermally averaged packets ($\overline{\theta}$).
The generalized persistence exponents
associated to $n$ crossings are also obtained. Specifying to
the process $A+A \to \emptyset$ or $A$ with probabilities
$(r,1-r)$, we compute exactly the exponents $\delta(r)$
and $\psi(r)$ characterizing the survival up to time $t$
of a domain without any merging or with mergings respectively,
and the exponents $\delta_A(r)$
and $\psi_A(r)$ characterizing the survival up to time $t$
of a particle $A$ without any coalescence or with coalescences 
respectively. $\overline{\theta}$, $\psi$ and $\delta$ obey
hypergeometric equations and are numerically surprisingly close to pure 
system exponents (though associated to a completely
different diffusion length). The effect of additional disorder in 
the reaction rates, as well as some open questions, 
are also discussed.

\end{abstract}


\widetext

\newpage



\section{Introduction} \label{sec:int}

\subsection{Overview}

Reaction diffusion processes are of wide interest in physics,
chemistry and biology \cite{privman}. In physics they present a
relatively simpler case of non equilibrium stochastic processes
with non trivial behaviour. Traditionally they have been
studied via mean field type methods (e.g. law of mass action, local 
chemical kinetics) \cite{smoluchovsky}.
However, in sufficiently low spatial dimension,
particle density fluctuations become dominant and mean field methods
become invalid \cite{ovchinnikov_zeldovich}.
The role of fluctuations in these processes has
thus been studied for a while, but has received renewed
attention recently \cite{kuzovkov_kotomin}, as new exact results in one dimension
\cite{privman}
and systematic renormalisation group studies have appeared
\cite{cardy_review}.
One interest of these models is their relation to phase
ordering kinetics via the "coarsening" of domain structures evolving
towards equilibrium \cite{bray}. In some cases, these can be seen
as reaction diffusion processes for defects, for instance domain walls
in one dimension or XY type vortices in two dimension, which
diffuse and can annihilate or coalesce upon meeting.
These coarsening processes have also been much studied recently,
especially in an effort to characterize their so-called
persistence (or survival or first passage) properties 
for single spins, domains or global magnetization
\cite{derrida_coarsening_phi4,krapivsky_benaim_redner,%
majumdar_sire,oerding_cornell_bray,derrida_hakim}

Although many results are now available for reaction diffusion
processes in homogeneous situations, comparatively little is known
on their dynamics in the presence of quenched disorder,
which is expected to play a role in many physical realisations.
It can be introduced in the models in several ways, e.g. in the reaction
rates or in the single particle diffusion. One can expect 
that it will strongly modify the behaviour of the system
in some cases by amplifying the role of spatial density
fluctuations. These effects are interesting, but 
difficult to study analytically because of the present lack of 
methods, beyond mean field approximations or
perturbation theory, to treat the dynamics of such
disordered systems.

Even in the absence of quenched disorder, there is an
apparently unlimited variety of behaviours in 
reaction diffusion systems. The more complex
ones, such as oscillatory or chaotic behaviours, become possible
for a large enough number of species 
\cite{velikanov_chaotic,nicolis_chaos,coullet_chaotic,coullet_chaotic_phd}.
In simpler cases, attempts have been made to identify possible 
universality classes and a wide class of models with
finite reaction rates, amenable to field theoretical treatments,
has been studied \cite{peliti_review,cardy_tauber}.
For instance, branching and annihilating random walks (BARW),
i.e reactions such as $A \to m A$ and $A+A \to 0$ or $A \to 0$
exhibit transitions from inactive (no particle) to 
active states, which were found to be
either in the universality class of directed percolation
\cite{grassberger_percodir,cardy_percodir,janssen_percodir}
(odd number of offsprings) or in the so called
parity conserving class (even number of offsprings)
\cite{cardy_tauber}.
This was confirmed by exact results in one dimension
\cite{mussawisade_oned}. Related types of models
describe epidemy propagation, such as 
$A + B \to 2 B$ (with rate $k$) and either 
$B \to A$ (recovery) or $B \to C$ (immune)
(rate $1/\tau$) were also studied via RG \cite{vanwijland_epidemy}
(see \cite{vanwijland_phd} for review).
The effect of quenched disorder has been studied
in this class of BARW models, via random rates
$k(x)$ and $\tau(x)$ but with limited success
as the RG flows to strong coupling \cite{vanwijland_phd}.
As for directed percolation with disorder, it is
still a largely open problem 
\cite{obhukov_disorder,janssen_disorder,moreira_disorder}.

There is a simpler class of homogeneous models without branching
(i.e without particle production),
such as $A+A \to 0$, $A+A \to A$
\cite{lee_annihilation,peliti_annihilation},
$A + B \to 0$ 
\cite{kang_redner,cheng_segregation}, etc..
which has still non trivial behaviour \cite{privman_short}.
One interesting phenomenon is that
in low enough dimension, the process becomes
diffusion limited rather than reaction limited.
Indeed particles in close proximity
react quickly and the remaining particles
are typically separated by a length related to
the pure diffusion length $L_0(t) \sim (D t)^{1/2}$.
This leads to a decay of specie density,
e.g. $n_A(t) \sim t^{-d/2}$ for $d<2$ in the case of
$A + A \to 0$, slower than the
mean field prediction $n_A(t) \sim t^{-1}$
valid for $d>2$
(for $A + B \to 0$ a related argument
leads $n_A(t) \sim t^{-d/4}$ for $d<4$) \cite{kang_redner}.
This type of results for such models in the pure case are
well established from heuristic arguments,
numerical simulations, perturbative RG 
\cite{lee_annihilation,peliti_annihilation,lee_cardy}
and in some cases rigorous methods
\cite{bramson_lebowitz,privman_short}.
It is now interesting to investigate how disorder will
modify these behaviours. With disorder, models in this
class are easier to study than the BARW type models,
although it is still a difficult task.
The reactions $A+A \to 0$ and $A+B \to 0$
have been studied using perturbative 
field theoretic RG methods for particles
diffusing in random flows, either in 
two dimension \cite{park_deem_rd98,park_deem_rd97} or
in a special hydrodymamic flow \cite{oerding_rd96}.
As can be expected from the study of single particle diffusion
in such flows \cite{fisher_friedan,ledou_lr},
the behaviour should be qualitatively different
in the case of {\it potential } disorder, which tends
to segregate the particles and slow the reaction (and the diffusion)
than for {\it hydrodynamic } flows which tend to mix the particles
and increase the effective reaction rate
(and leads to hyperdiffusion). The competition which
arises when both components are present
has been studied very recently in $d=2$ in \cite{park_deem_rd98}
and in $d=2 -\epsilon$ \cite{richardson_cardy99}.
Remarkably, the one dimensional problem seems
quite far from the reach of such perturbative RG methods
and no generic result is available 
at present \cite{footnote4} 
in that case, hence the interest of the 
present study. Indeed, in $d=1$ only potential disorder
can exist and is known to lead to ultraslow single particle diffusion
described by a strong disorder (i.e zero temperature) fixed point
\cite{footnoterg}.
To make progress in $d=1$ requires developing non perturbative
techniques, which is the aim of the present work. 

\subsection{Model}
\label{secmodel}

In this paper we study a broad class of reaction diffusion models
where particles diffuse on a one dimensional lattice
and can react or annihilate upon meeting. Apart from 
their reactions the particles are non interacting.
More specifically, each site of the lattice can be
in one of several possible ``states'',
labeled $k=0,1,..n-1$. $k=0$ corresponds to the empty state
with no particle present at that site. $k=1,..,n-1$ correspond
to the presence of particles of different types.
When two particles (i.e states) $k_1>0$ and $k_2>0$ meet they react
and give a another state $k$ with a probability $W^{k}_{k_1,k_2}$.
$k$ may be the empty state $k=0$, corresponding to
an annihilation. The reaction is thus a stochastic process 
\begin{eqnarray}
k_1 + k_2 \to k ~~~~ \hbox{with probability} ~~~ W^{k}_{k_1,k_2} 
\label{general}
\end{eqnarray}
characterized
by a fixed transition probability matrix which satisfies 
\begin{eqnarray}
\sum_{k} W^{k}_{k_1,k_2}=1
\label{matrixW}
\end{eqnarray}
The matrix $W^{k}_{k_1,k_2}$ can be extended to include
$k_1=0$ by defining
\begin{eqnarray}
W^{k}_{0,k'} = W^{k}_{k',0} = \delta_{k,k'}
\label{matrixW}
\end{eqnarray}
for any $k,k'$, which is the property expected
for an empty state ($A + 0 \to A$ with probability 
1).

One prominent example will
be identical particles $A$ which react upon meeting as
\begin{eqnarray} 
&& A + A \to \emptyset ~~~~ \hbox{with probability} ~~~ r \nonumber \\
&& A + A \to A ~~~~ \hbox{with probability} ~~~ 1-r
\label{potts}
\end{eqnarray}
In this case there are only two states: $k=0$ corresponds to
no particule present ($\emptyset$) and $k=1$ to
one particule present $A$. The transition matrix is
then $W^0_{0,0}=W^0_{0,1}=W^0_{1,0}=W^1_{0,0}=0$, 
$W^1_{0,1}=W^1_{1,0}=1$ and $W^0_{1,1}=r, W^1_{1,1}=1-r$.

We will obtain results for processes within
the above class (\ref{general}) and study 
some specific examples. We will restrict 
to symmetric reaction rates $W^k_{k_1,k_2}=W^k_{k_2,k_1}$.
Asymmetric rates, depending on the side from
which the two species come in contact, can be defined in 
$d=1$ and can be studied by the same methods.
We will mostly consider reaction
diffusion processes with only a {\it finite} number of possible
states (or species). Processes with unbounded number of states
($n = \infty$) can also be studied by the present method, and
we will give some examples. Classifying the
much larger variety of complex behavior possible in that
case is beyond the scope of the present study.
Other extensions include randomness in the reaction
rates, which we will briefly discuss in the end.

Up to now we have not specified the way in which 
particle diffuse, nor the reaction rates. Let us
first concentrate on the process (\ref{potts}) and recall the
known results in the case of pure diffusion (i.e homogeneous hopping rates),
which has been extensively studied. 
It is of particular interest in one dimension since 
it is also a model for zero temperature 
domain growth in the ferromagnetic $q$-states Potts model (with
Glauber dynamics) where $r=1/(q-1)$ \cite{derrida_hakim}. 
The case $q=2$ (Ising) corresponds to walkers (i.e domain walls)
always annihilating when they meet and $q=\infty$ to walkers
always coagulating \cite{footnote_hidden}. It is known that
the reaction rate can be chosen infinite (immediate reaction 
upon meeting) without changing the universality class, and the
same will hold here in presence of disorder, hence our general
choice of model (\ref{general}). For all $r$ 
the concentration of particles $A$ is known to decrease as:
\begin{eqnarray} 
n_A(t) \approx  c(r) (D t)^{- 1/2} 
\end{eqnarray}
where the ($r-dependent$) coefficient
is expected to be universal (e.g $c(1)=(8 \pi)^{-1/2}$,
\cite{lushnikov}). More detailed properties, such as persistence,
have also been studied.
The probability $S(t)$ that no particle
$A$ (domain wall ) has crossed a given point $O$ up to time
$t$ has been shown to decay as:
\begin{eqnarray} 
S(t) \sim L_0(t)^{- \theta(r)} 
\end{eqnarray}
where $L_0(t) \sim \sqrt{t}$ is the characteristic length
and $\theta(r)$ the so-called persistence exponent.
$S(t)$ also corresponds to the probability that a
spin has never been flipped up to time
$t$ in the Potts model. The exact expression of
$\theta(r)$ was obtained in \cite{derrida_hakim,derrida_hakim_long}.
The domain size distribution has also been computed
for this process in \cite{derrida_zeitak}.
Two new independent exponents, $\psi$ and $\delta$
were introduced and studied in
\cite{krapivsky_benaim,krapivsky_redner} to characterize
the persistence (survival) of domains for this
model, but up to now no exact result is available
for these exponents. The concept of persistence
properties was extended to other observables,
finite temperature and studied in a variety
of other models: persistence for global order
parameter in \cite{oerding_cornell_bray},
spin block persistence 
in \cite{sire_cueille},
generalized persistence and large deviations
in \cite{dornic_godreche,howard_godreche} and
persistence for fluctuating interfaces
\cite{krug_persistence}.

In this paper we study the case where the hopping rates
are inhomogeneous with short range correlations,
corresponding to random local bias. The generic model 
for this type of disorder
is the Sinai model where each particle performs
Arrhenius diffusion in the same energy landscape
$U_n$ where the local random forces $U_{n}-U_{n+1}$
are independent random variables, of zero average
(we restrict the present analysis to zero global bias).
Various analytical results are known for the single particle Sinai model
\cite{sinai,kesten1,kesten,derrida_pomeau,ledou_1d,monthus_diffusion,%
laloux_pld_sinai}. Diffusion is ultraslow
as $x \sim L(t)=(\ln t)^2$. Recently we have reexamined
this model \cite{us_prl,us_long} using a real space renormalization group
method (RSRG) which yields exact results at long time. In the present paper
we apply the RSRG method to study reaction diffusion of the type
(\ref{general}) for particles in a Sinai landscape.
Some of the results have already appeared in \cite{us_prl}.
Although we give here a detailed treatment of the
reaction diffusion RSRG, we will rely on Ref.
\cite{us_long} for all details concerning the
single particle diffusion aspects of the problem
(which we will only sketch, referring the reader 
to \cite{us_long} for details). Note that we 
consider here only models where all particles 
share the same diffusion property (i.e. see the same landscape,
and have same diffusion coefficient). Thus this does not
include reaction diffusion models such as Ising domain walls in
a random field, for which a specific treatment is necessary and
which are studied in \cite{us_prl,us_long_rf}.
Similarly, relations to other problems such as disordered quantum spin
chains \cite{ma_dasgupta,danfisher_rg1,danfisher_rg2,monthus_golinelli}
or disordered free fermion models are discussed in
\cite{us_prl,us_long}. In particular we have chosen to discuss our present
results exclusively in terms of reaction diffusion dynamics,
and not in their equivalent formulation as (non hermitian) 
1D quantum models (see \cite{quantum} for details
of such relations in the pure case). Finally, note that an exact RG has also been 
applied to the problem of coarsening of the pure 1D $\Phi^4$
model at zero temperature for which persistence exponents 
have been computed \cite{derrida_coarsening_phi4,majumdar_bray}.

As for the single particle problem, the RSRG 
method allows us to compute a number of
quantities, and, remarkably, even some which
are not known for the corresponding pure
model (e.g. the domain persistence exponents $\delta$ and $\psi$).
This makes the disordered case all the more
interesting to study. We find that 
reaction diffusion processes in a Sinai landscape are strongly
controlled by the ultraslow diffusion, e.g
the relevant length scale is the 
diffusion length $L(t) \sim (\ln t)^2$, but that they still
possess non trivial reaction properties. We will
characterize a broad set of universality classes, containing all
reactions of type (\ref{general}). The reaction times (provided they are finite)
do not affect any universal quantity, so that we
can consider the reactions as instantaneous
for practical purpose. As discussed in
\cite{us_long} there are other single particle
diffusion models with short range correlated disorder
in one dimension apart from Sinai's model universality class,
such as random barriers (symmetric hopping rates) or random wells.
For interesting behaviour to occur, however, algebraically
broad distributions are required from the start.
Some results for reaction diffusion processes with
this type of single particle diffusion have been obtained in
\cite{schutz_oned}. 

The outline of the paper is as follows. In Section \ref{rgmethod}
we detail the RSRG method, first recalling known results 
in the case of a single particle in \ref{rgsinai}, then
deriving the RSRG equation for reaction diffusion models
\ref{rgreacdiff}. The fixed points of this equation,
and some physical properties of the corresponding asymptotic states
are studied in \ref{fixedpoints} and \ref{physical} respectively.
Section \ref{asymptoticdynamics} is devoted to a detailed
analysis of the dynamics near attractive or repulsive fixed points
and of the convergence towards the asymptotic states. Throughout the paper 
we apply our results to the process (\ref{potts}) but in
Section \ref{examples} we discuss some applications 
to other examples of processes.
In Section \ref{persistence} we study persistence properties.
Section \ref{conclusion} contains the conclusion.
Some more technical but useful details are contained 
in the appendices.

\section{RSRG method for reaction diffusion and asymptotic states}

\label{rgmethod}

\subsection{RSRG for Sinai landscape and single particle diffusion}

\label{rgsinai}

\begin{figure}[thb]

\centerline{\fig{8cm}{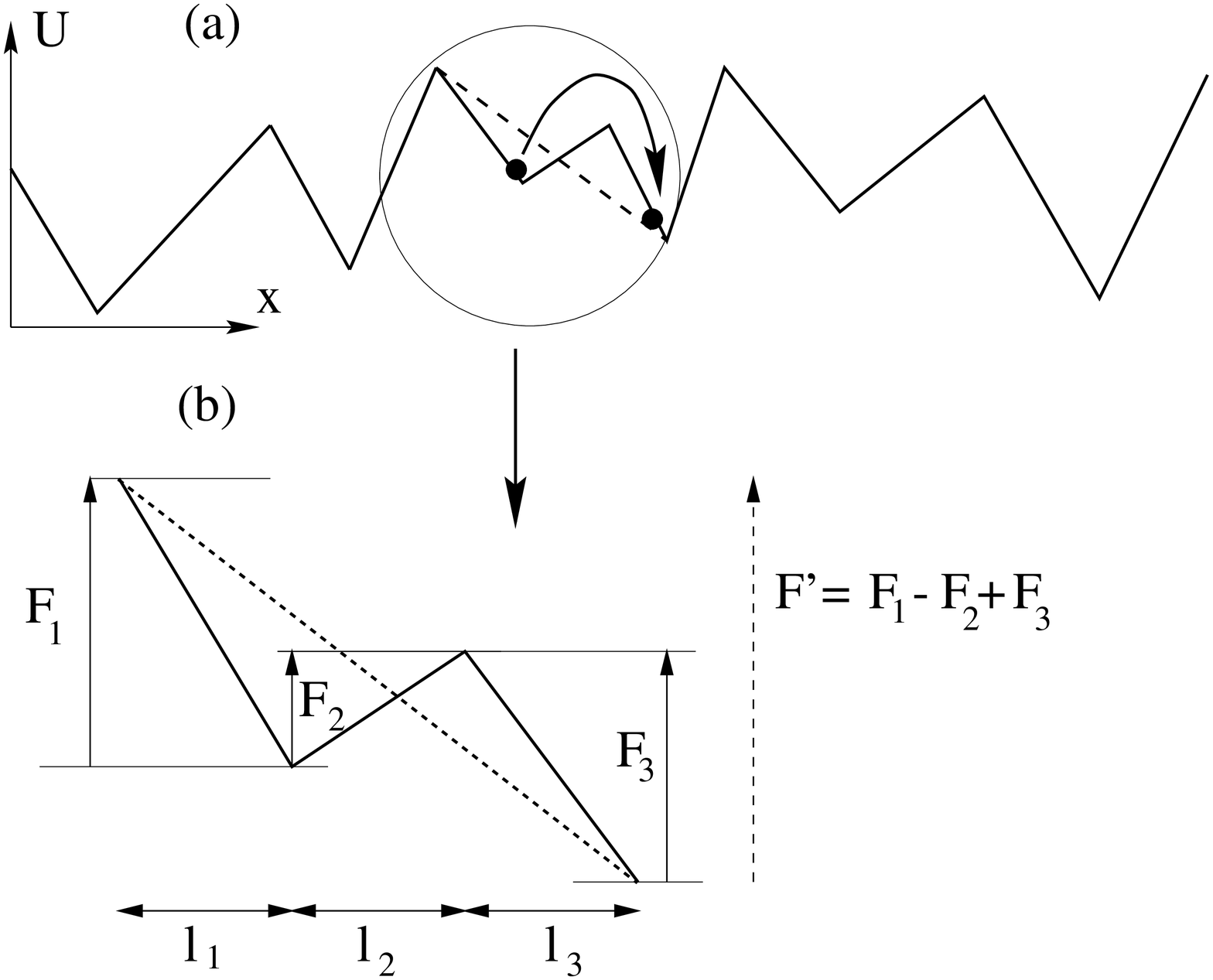}} 
\caption{\narrowtext (a) Energy landscape in the Sinai model (b) 
decimation method: the bond with the smallest barrier
$F_{min}=F_2$ is eliminated, resulting in three bonds being grouped into
a single one. \label{fig1} } 

\end{figure}

The model for the diffusion of a single particle
in one dimension can be defined, with no loss of generality \cite{us_prl},
as the Arrhenius diffusion in a "zigzag" potential $U(x)$
represented in Fig. \ref{fig1} (a). It consists in a set
of bonds, each bond (between $x_i$ and $x_{i+1}$)
being characterized by an energy barrier $F_i=|U_i - U_{i+1}|$ (where $U_i=U(x_i)$)
and a length $l_i=|x_{i+1} - x_i|$. The energy landscape is chosen
by choosing a pair of bond variables $F,l$, independently
from bond to bond, from a distribution $P(F,l)$ normalized to unity.

The RG procedure, which captures the long time behaviour in a given energy
landscape, is illustrated in
Fig. \ref{fig1} (b) and consists in the iterative decimation of 
the bond with the {\it smallest barrier} \cite{us_prl},
say $F_2$, and to replace the three bonds $1,2,3$
by a single renormalized bond with 
barrier 
$F'=F_1 - F_2 + F_3$ and length $l'=l_1 + l_2 + l_3$.
The new variables remain {\it independent} 
from bond to bond. To write the corresponding RG equation 
it is convenient to introduce $\Gamma$ as the smallest
remaining barrier at a given stage of 
the decimation and the rescaled variables
$\eta = (F-\Gamma)/\Gamma$ and 
$\lambda= l/\Gamma^2$.
The RG equation 
for the probability distribution \cite{footgamma}
$P^\Gamma(\eta,\lambda)$ reads \cite{us_prl}:
\begin{eqnarray}   \label{rgbond}
(\Gamma \partial_\Gamma - (1+\eta) \partial_\eta - 2 \lambda \partial_\lambda
- 3) P^\Gamma(\eta,\lambda) = 
P^\Gamma(0,.)*_\lambda P^\Gamma(.,.) *_{\eta, \lambda} P^\Gamma(.,.)
\end{eqnarray}
and coincides with the one derived in \cite{danfisher_rg1}
for the closely related problem of disordered quantum spin chains.
The symbol $*_\lambda$ denotes a convolution with respect
to $\lambda$ only and $*_{\eta,\lambda}$ 
with respect to both $\eta$ and $\lambda$. 
The probability distribution is normalized
to unity as $\int_0^{+\infty} d\eta \int_0^{+\infty} d\lambda
P^\Gamma(\eta,\lambda) = 1$.

The landscape is characterized by 
the large scale variance of the potential:
\begin{eqnarray}
\overline{ (U_i-U_j)^2 }
\approx  2 \sigma |l_{i-j}|
\end{eqnarray}
with $l_{i-j}$ the distance from site i to site j,
which is exactly preserved by the RG. Thus we will set
$\sigma=1$ in the following. Restoring $\sigma$ simply amounts to
a rescaling of lengths, and in particular 
$\sigma$ drops out of all (universal) ratios of lengths 
that we consider later. As shown in \cite{danfisher_rg1,danfisher_rg2},
the RG equation (\ref{rgbond}) leads
at large $\Gamma$ (using Laplace transforms)
to the following fixed point $P^*(\eta,\lambda)$ :
\begin{eqnarray}
P^*(\eta,\lambda) = LT_{s \to \lambda}^{-1}
\left(  \frac{\sqrt{s}}{\sinh \sqrt{s}  } \ 
e^{-\eta\sqrt{s}\coth \sqrt{s} }  \right)
\label{solu}
\end{eqnarray} 
Thus for large $\Gamma$ 
one finds that the average bond length $\overline{l}_{\Gamma}$
and the number of bonds $n_\Gamma$ per unit length
are \cite{footnotel}, respectively:
\begin{eqnarray} \label{scaling}
\overline{l}_{\Gamma} =  \frac{1}{2}\Gamma^2 \qquad n_\Gamma = \frac{2}{\Gamma^2}
\end{eqnarray}

The renormalized landscape allows to study the dynamics
of a single walker starting from a given point $O$
at $t=0$. The decimation of barriers smaller than
\begin{eqnarray}
\Gamma = T \ln t
\end{eqnarray} 
corresponds to the elimination of (logarithmic)
time scales shorter than the Arrhenius time $t$
for the particle to cross the barrier. We are 
choosing everywhere time units such that
the (non universal) microscopic attempt time scale $t_0$ be set
to unity (arbitrary units can be recovered by 
setting $\Gamma = T \ln(t/t_0)$ in what follows \cite{footnotet0}).
Since at long time (i.e large $\Gamma$)
the renormalized landscape consists entirely of deep valleys
separated by high barriers,
a good approximation to the long time dynamics is to place 
the walker {\it at the bottom} of the renormalized valley at scale
$\Gamma = T \ln t$
which contains the starting point $O$,
since with high probability it will be near to that point \cite{sinai}.
Upon proper rescaling of space and time this approximation 
becomes in fact {\it exact} as $\Gamma$ tends to $+ \infty$.
This defines what we will call
the ``effective dynamics'' in the following and is illustrated 
in Fig. \ref{fig1}. This allows to recover the scaling
$x \sim (\ln t)^2$ for the single particle diffusion
as well as many other exact results detailed
in \cite{us_prl,us_long}. Since it is customary,
when studying reaction diffusion processes, to
compare densities of reactants with a characteristic 
scale of diffusion, we give here the the exact expression 
for the single particle
r.m.s displacement, or ''diffusion length''
at large time:
\begin{eqnarray}
\sqrt{\overline{< x^2(t) >} } \approx
\frac{1}{6} \sqrt{ \frac{61}{5} } T^2 (\ln t)^2
\end{eqnarray} 

To study reaction diffusion processes it will be
necessary to consider ``valleys'' (two consecutive bonds
sharing a common potential minimum).
We thus slightly generalize the above RG
equation (\ref{rgbond}) to follow the distribution
of renormalized valleys. The RG equation for the valley probability
distribution $P^{\Gamma}(\eta_1,\eta_2)$ at scale $\Gamma$ in rescaled
variables $(\eta_1,\eta_2)$ reads
\begin{eqnarray}
&& \Gamma \partial_\Gamma P^{\Gamma}(\eta_1,\eta_2)
=\left[(1+\eta_1) \partial_{\eta_1} + 
(1+\eta_2) \partial_{\eta_2} ) + 2 \right] 
P^{\Gamma}(\eta_1,\eta_2) \nonumber \\
&& + P^{\Gamma}(\eta_1,.) *_{\eta_2} P^{\Gamma}(0, .) + 
P^{\Gamma}(.,0) *_{\eta_1} P^{\Gamma}(., \eta_2)
\label{rgvalley}
\end{eqnarray}
where we have omitted (i.e integrated over) the lengths
for simplicity. The large time ($\Gamma$) behaviour
of this equation can be studied similarly.
Valley distributions which have the 
decoupled form $P^{\Gamma}(\eta_1,\eta_2) =
P^{\Gamma}(\eta_1) P^{\Gamma}(\eta_2)$ 
where $P^{\Gamma}(\eta)$ satisfies the bond RG equation (\ref{rgbond})
are of course solution of the RG equation for
valleys (\ref{rgvalley}). The subspace of such
decoupled distributions (called decoupled subspace
in the following) is thus preserved by RG.
Since the initial condition is uncorrelated,
the RG flow defined by (\ref{rgvalley}) remains in this
decoupled subspace, and converges towards the fixed point
$P^*(\eta_1,\eta_2)= P^*(\eta_1) P^*(\eta_2) =e^{-\eta_1-\eta_2}$.
This convergence result extends to the case of small
correlations between valley sides as will be discussed below.

It was shown in \cite{danfisher_rg2} that the convergence towards the
bond fixed point $P^*(\eta)=e^{-\eta}$ is like $\frac{1}{\Gamma}$
with eigenvector $(1-\eta) e^{-\eta}$. Thus 
the convergence towards the the valley fixed point
$P^*(\eta_1,\eta_2)=e^{-\eta_1-\eta_2}$ within the
decoupled subspace is of the form:
\begin{eqnarray}
\left( e^{-\eta_1} + \frac{c}{\Gamma} (1-\eta_1) e^{-\eta_1} +...\right)
\left( e^{-\eta_2} + \frac{c}{\Gamma} (1-\eta_2) e^{-\eta_2} +...\right)
= e^{-\eta_1-\eta_2} (1 +\frac{c}{\Gamma} (2-\eta_1-\eta_2) ) +...
\end{eqnarray}
and is also as $\frac{1}{\Gamma}$
with eigenvector $(2-\eta_1-\eta_2) e^{-\eta_1-\eta_2}$.

\subsection{RG equations for reaction-diffusion}

\label{rgreacdiff}

We now turn to diffusion reaction models of type
(\ref{general}) in 
one dimensional landscapes with random local biases.

\begin{figure}[thb]

\centerline{\fig{12cm}{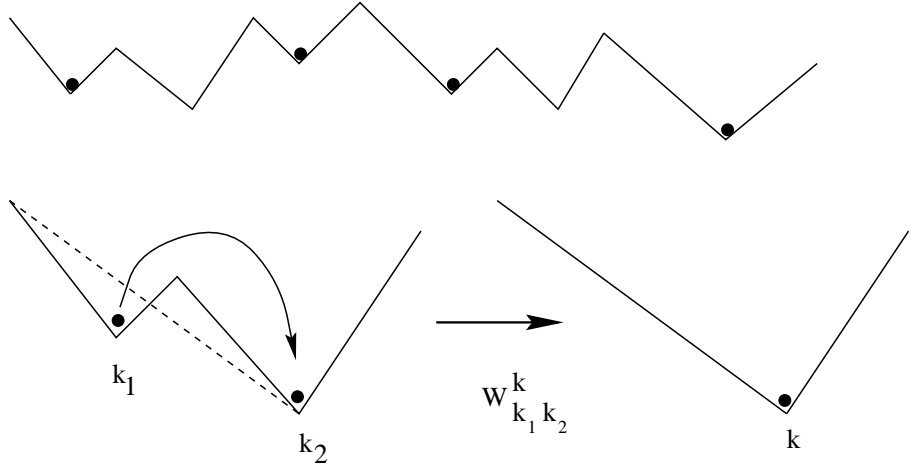}} 
\caption{\narrowtext (a) macroscopic state of the reaction
diffusion process: each renormalized valley is either empty
($k=0$) or contains a particle of type $k>0$ (b)
at time scale $t$ such that the barrier $F=\Gamma = T \ln t$
is decimated, the state (particle) $k_1$ in the decimated
valley moves to the neighboring valley and reacts
with $k_2$ to produce $k$ with probability $W^k_{k_1,k_2}$,
as the two valleys are merged into a single renormalized one 
containing $k$ \label{fig2}} 

\end{figure}

From the results on the dynamics of a
single particle in Sinai landscape recalled in the
preceding section, it is clear that one can study most of
the properties of the initial reaction diffusion problem
by following its evolution under the effective dynamics.
It also becomes obvious that one must now consider valleys,
and the species contained in these valleys.
At the decimation time scale $\Gamma = T \ln t$
in some places in the system, two valleys will merge into
one and the reaction (\ref{general})
governed by the rates matrix $W$
will take place. This process is 
illustrated in Fig. \ref{fig2}.
The errors made by this approximation are
expected to become
again smaller at large time, as will be discussed
later on.

The general method to study the process (\ref{general})
is thus to associate to each valley the specie which it contains,
which is one of several possible states $k$ ($k=0$ being
the empty state). A convenient initial model is thus one where 
each valley and its content is statistically independent
and characterized by a probability distribution
$P^{\Gamma}_k(\eta_1,\eta_2)$ with 
$\sum_k P^{\Gamma}_k(\eta_1,\eta_2)=P^{\Gamma}(\eta_1,\eta_2)$. It remains
so under the RG. The effective dynamics is described by the RG equation:

\begin{eqnarray}
&& \Gamma \partial_\Gamma P^{\Gamma}_k(\eta_1,\eta_2)
=((1+\eta_1) \partial_{\eta_1} + (1+\eta_2) \partial_{\eta_2} ) 
+ 2) P^{\Gamma}_k(\eta_1,\eta_2) \nonumber \\
&& + W^k_{k_1,k_2} [ P^{\Gamma}_{k_1}(\eta_1,.) *_{\eta_2} P^{\Gamma}_{k_2}(0, .) 
+ P^{\Gamma}_{k_1}(.,0) *_{\eta_1} P^{\Gamma}_{k_2}(., \eta_2) ]
\label{rgreadiff}
\end{eqnarray}
where summation over repeated indices is implied.
The summation over $k$ yields back the valley RG equation (\ref{rgvalley}).
Since the average length of a valley is $2 \overline{l}_\Gamma$,
The total concentration $n_k(t)$ of a given specie $k$ 
(the total number of $k$-particles per unit of length) is 
given as:
\begin{eqnarray}
n_k(t) = \frac{1}{2} n_\Gamma p_k^\Gamma
\end{eqnarray}
where:
\begin{eqnarray}
p_k^\Gamma = \int_{\eta_1,\eta_2} P^\Gamma_k(\eta_1,\eta_2)
\end{eqnarray}
and $n_\Gamma$ is the number of remaining bonds (\ref{scaling})
per unit length at scale $\Gamma = T \ln t$. We stress that the RG equation
(\ref{rgreadiff}) is more complicated to analyze than
(\ref{rgvalley}) since it cannot in general be factorized
into bond distributions.

However, it turns out that there is still a simple subspace of distributions
which is exactly preserved by the RG equation (\ref{rgreadiff}).
It is the subspace of functions of the sum $\eta_1+\eta_2$:
\begin{eqnarray}
P^{\Gamma}_k(\eta_1,\eta_2)=H^{\Gamma}_k(\eta=\eta_1+\eta_2)
\end{eqnarray}
where the functions $H_k(\eta)$ satisfies:
\begin{eqnarray}
 \Gamma \partial_\Gamma H^{\Gamma}_k(\eta)
=((2+\eta) \partial_{\eta} + 2) H^{\Gamma}_k(\eta) 
 + W^k_{k_1,k_2}  H^{\Gamma}_{k_1}(.) *_{\eta} H^{\Gamma}_{k_2}( .) 
\end{eqnarray}
which conserved the normalization 
$\sum_k p_k^{\Gamma}=\int_0^{\infty} 
d \eta \ \eta \sum_k H^{\Gamma}_k(\eta)$. This subspace plays
an important role in the following. Already one sees that
both the fixed point $e^{-\eta_1-\eta_2}$ and the
leading eigenvector $(2-\eta_1-\eta_2) e^{-\eta_1-\eta_2}$
of the linearized landscape RG equation (\ref{rgvalley})
belongs to this subspace.

\subsection{Fixed points of the RG equations and asymptotic states}

\label{fixedpoints}

We now determine the fixed point solutions of the 
RG equation for the valley distributions (\ref{rgreadiff}).
We already know that the sum $\sum_k P^\Gamma_k(\eta_1,\eta_2)
=P^\Gamma(\eta_1,\eta_2)$ converges towards the 
fixed point of equation (\ref{rgvalley})
$P^*(\eta_1,\eta_2)=e^{- (\eta_1 + \eta_2)}$ 
which describes the landscape. It is thus natural to look
for fixed points of (\ref{rgreadiff}) of the following form
\begin{eqnarray}
P^*_k(\eta_1,\eta_2) = p^*_k ~ P^*(\eta_1,\eta_2)= p^*_k e^{- (\eta_1 + \eta_2)}
\label{fixedpointeta}
\end{eqnarray}
where $p_k^* \ge 0$ and $\sum_k p^*_k =1$ by normalization.
Plugging this form into (\ref{rgreadiff}) leads to
a consistent $\Gamma$ independent solution if the $p_k^*$
satisfy the condition:
\begin{eqnarray}  \label{eq}
p^*_k = W^{k}_{k_1,k_2} p^*_{k_1} p^*_{k_2} 
\end{eqnarray}
Note that any solution of this equation satisfies a priori $\sum_k p^*_k=0$ or $1$,
as a consequence of (\ref{matrixW}). Thus apart from the
unphysical solution of (\ref{eq}) where all $p_k^*$ vanish, all other solutions 
are automatically correctly normalized.

In general the equation (\ref{eq}) has several solutions and thus
there are several fixed points to the valley equations (\ref{rgreadiff}).
Clearly some of these fixed points are attractive and correspond
to possible large time asymptotic states for the reaction diffusion
process while other fixed points are repulsive.
In some cases several attractive fixed points can coexist
and lead to a non trivial phase diagram.

The stability of each fixed point, as well as their convergence 
property, will be studied in detail in the next Section.
Here we just mention one important result.
The dynamics in the vicinity of a fixed point
$p_k^*$ is determined by the following stability matrix:

\begin{eqnarray}  \label{matrixM}
M_{k j} = W^{k}_{k',j} p^*_{k'}
\end{eqnarray}
We denote by $\mu_{\alpha}$ the eigenvalues
of $M$, and by $p^{\alpha}_j$ the associated eigenvectors
$  M_{k j} p^{\alpha}_j = \mu_{\alpha} p^{\alpha}_k$.
For $n-1$ reacting species (in addition to the empty state $k=0$),
$M$ is an $n$-dimensional matrix
with non-negative elements. One eigenvalue is trivial, 
being simply $\mu_1=1$ of eigenvector 
$p^{\alpha=1}_k \propto p^*_k$ (\ref{eq}).
The other $n-1$ eigenvalues $\mu_\alpha$ ($\alpha=2,...n$),
which can be complex in general,
have a smaller modulus, due to the Perron-Froebenius theorem.
We will focus in the following
on simple enough processes where all $\mu_\alpha$ are real,
but we will also give example of more general reaction
diffusion where complex eigenvalues arise (e.g.
reactions with cycles).

Let us consider here only fixed points with
all $\mu_\alpha$ being real. 
The result of the next Section is that, in that case,
a given fixed point if
attractive is the $n-1$ eigenvalues $\mu_{\alpha} < 1/2$
for $\alpha=2,...n$. It is
repulsive if at least one of these $\mu_{\alpha}> 1/2$
(and is then repulsive along the corresponding eigendirection).

For example, in the case of
the reaction-diffusion process (\ref{potts}),
there are two solutions of (\ref{eq}) and thus two fixed
points. One is the empty state
E=($p^*_0=1$ , $p^*_A=0$) and the other is S=($p_0^*=r/(1+r)$, $p_A^*=1/(1+r)$).
The stability matrix associated to E is simply the 
$2 \times 2$ identity matrix (i.e. $\mu_1=1=\mu_2$)
and this fixed point is thus repulsive. 
The matrix associated to the state S 
reads $M_S=( (\frac{r}{r+1}, \frac{r}{r+1}),
(\frac{1}{r+1}, \frac{1}{r+1}))$ with eigenvalues $\mu_1=1$ and $\mu_2=0$.
The fixed point S is thus attractive and corresponds to the
asymptotic state which represents the large time behaviour of
the system.

In fact the reaction-diffusion process (\ref{potts}) possesses
an interesting property: the outcome of a sequence of reactions
does not depend on the order it was performed. We call these
processes ``associative processes''. They have the property that
$M^2 = M$, i.e $\mu=0,1$. Some properties of these
associative processes are detailed in Appendix (\ref{appassoc}).

\subsection{physical properties of the asymptotic states}

\label{physical}

We now study some physical consequences. Each attractive
fixed point corresponds to a possible large time behaviour of
the system, i.e an asymptotic state.
If there are several attractive fixed point,
the one chosen by the system will depend on the initial
value of the parameters (mainly the specie concentrations).

From the results (\ref{fixedpointeta}) and (\ref{scaling})
we obtain that in an asymptotic state (characterized by a
set of $p^*_k$ solution of (\ref{eq}), the
density of specie $k$ behaves at large time as:

\begin{eqnarray}
n_k(t) = \frac{1}{2} n_\Gamma p^*_k = \frac{p^*_k}{T^2 \ln^2(t/t_0)}
\label{density}
\end{eqnarray}
where we have restored the microscopic attempt time scale $t_0$
\cite{footnotet1}. 
Note that this result (\ref{density}) represents the leading large time
contribution, subleading corrections (which become dominant only 
if $p^*_k=0$) will be determined
in the next section. Interestingly, this leading behaviour is
independent of the initial concentration (provided
it is in the basin of attraction of the fixed point).
This universality property can be further characterized by
computing universal amplitudes. In pure models
a commonly studied amplitude is the
product of specie concentration by the diffusion volume.
In a disordered model one has more choices of definitions,
but we will define an amplitude as in 
\cite{richardson_cardy99}. Here we find the exact result
for the following {\it universal amplitude} (associated to specie $k$):

\begin{eqnarray}
 {\cal A}_k \equiv \lim_{t \to \infty} n_k(t) \sqrt{\overline{\langle x^2(t)  \rangle}}
= p^*_k \frac{1}{6} \sqrt{\frac{61}{5}} 
\end{eqnarray}
This gives for instance ${\cal A}(r) = 1/(1+r) \frac{1}{6} \sqrt{\frac{61}{5}}$
for the process (\ref{potts}), i.e 
${\cal A}= 0.291071..$ for $A + A \to 0$ studied in 
\cite{richardson_cardy99} by perturbative methods
\cite{footnotecardy}.

From the statistical independence of valleys in an
asymptotic state, information about the spatial
distribution of the species can also be obtained.
For instance one can define ``domains'', in the simplest
case as intervals between particles
(i.e between non empty states $k \neq 0$, irrespective of their content)
We can now compute exactly the distribution of
size of ''domains''. From the above form of the fixed points
the normalized distribution $D_\Gamma(l)$ of domain sizes $l$
in an asymptotic
state takes a scaling form $D_\Gamma(l) = \frac{1}{\Gamma^2}
D^*(\lambda=l/{\Gamma}^2)$ which can be computed as follows.

The above RG equation for valley distributions (\ref{rgreadiff}) can be
readily extended to $P_k(\eta_1,\eta_2,\lambda_1,\lambda_2)$
which takes into account the rescaled lengths $\lambda_1=l_1/\Gamma^2$,
$\lambda_2=l_2/\Gamma^2$ of the two bonds of the valley, extending (\ref{rgbond}).
The generalized fixed point (\ref{fixedpointeta}) reads:
\begin{eqnarray}
P^*_k(\eta_1,\eta_2,\lambda_1,\lambda_2) 
= p^*_k P^*(\eta_1,\lambda_1) P^*(\eta_2,\lambda_2)
\label{fixedpointeta2}
\end{eqnarray}
where $P^*(\eta,\lambda)$ is the fixed point solution (\ref{solu})
of the bond RG equation.
A domain as defined above, is thus a set of consecutive
empty valleys between two occupied valleys,
together with one bond in each of the occupied valleys
(see Fig. \ref{fig2}). Since in the asymptotic state valleys
are statistically independent and either empty ($k=0$) with probability
$p_0^* P^*(\lambda_1) P^*(\lambda_2)$ 
(where $P^*(\lambda) = \int_\eta P^*(\lambda)$)
or contains a particle ($k \neq 0$) with probability
$(1- p_0^*) P^*(\lambda_1) P^*(\lambda_2)$,
one easily obtains the Laplace transform of
$D^*(\lambda)$ as:
\begin{eqnarray}
\hat{D^*}_{p_0^*}(s) = \int_0^{+\infty} d\lambda e^{- s \lambda} D^*(\lambda)
= \frac{(1-p_0^*) P^*(s)^2}{(1-p_0^* P^*(s)^2)} =
\frac{1-p_0^*}{\cosh^2{\sqrt{s}} -p_0^*}
\label{lengths}
\end{eqnarray}
where we have used the explicit form (\ref{solu}) 
for the fixed point bond distribution. Formula
(\ref{lengths}) can be inverted and yields
the distribution of rescaled domain sizes:
\begin{eqnarray}
&& D^*_{p_0^*}(\lambda) = \tan \alpha \sum_{n=-\infty}^{+\infty}
(\alpha + n \pi) e^{- \lambda (\alpha + n \pi)^2}  \\
&& = \frac{\tan \alpha}{ \sqrt{\pi} \lambda^{3/2} }
\sum_{m=-\infty}^{+\infty} m \sin (2 \alpha m)  e^{- \frac{m^2}{\lambda} } 
\label{distriblengths}
\end{eqnarray}
with $\alpha =\arccos{p_0^*}$.

This can be applied, e.g. in the case of the
process (\ref{potts}). Substituting $p_0^* = r/(1+r)$
in (\ref{distriblengths}) yields the distribution
of distances between neighboring walkers A.
Note that the case where walkers A always 
coalesce upon meeting ($r=0$, $p_0=0$)
corresponds to $\alpha \to \frac{\pi}{2}$ and in this 
limit (\ref{distriblengths}) becomes:
\begin{eqnarray}
&& D_{p_0^*=0}^*(\lambda) = \sum_{n=-\infty}^{+\infty}
(2 \lambda \pi^2 (n+1/2)^2 - 1) 
e^{- \lambda \pi^2 (1/2+ n )^2}  \\
&& = \frac{2}{ \sqrt{\pi} \lambda^{3/2} }
\sum_{m=-\infty}^{+\infty} (-1)^{m+1} m^2  e^{- \frac{m^2}{\lambda} } 
\end{eqnarray}

It is interesting to compare the result (\ref{distriblengths}) 
concerning the disordered case with the result
of Derrida and Zeitak \cite{derrida_zeitak} for the
case of homogeneous hopping rates. For small domain sizes ($\lambda \to 0$),
 the distribution vanishes
much faster in the disordered
case (as $\sim \lambda^{-3/2} \exp(-1/\lambda)$) than in the
pure case ( as $\sim \lambda$).
For large domain sizes ($\lambda \to \infty$),
 both have exponentially decaying behaviour
(except for $q=+\infty$, i.e $r=0$
in the pure case). In addition, in the present
case the consecutive domains lengths are statistically
independent, which is not the case for the pure system.

The above calculation is easily generalized to compute the
distribution of relative distances between two walkers of
a given species $k$, simply by substituting $p_0^* \to 1-p_k^*$
in the above formula (\ref{distriblengths}).

\section{Dynamics near fixed points and asymptotic states}

\label{asymptoticdynamics}

In this Section we study the dynamics near the
possible fixed points of the valley RG equation
(attractive and repulsive). 

We will first focus strictly
on the effective dynamics exactly described by
the RG equation (\ref{rgreadiff}), and mention
some possible corrections in the real dynamics 
at the end of the section.

For the effective dynamics we will solve the problem 
in two steps. As mentionned above the matrix $M$ in (\ref{matrixM})
and its eigenvalues $\mu_\alpha$ control the asymptotic dynamics.
Interestingly they readily provide an approximation of the dynamics,
which we will call the ``uniform approximation'', which is interesting
as it allows to classify the spectrum of eigenperturbations
and, in the case of real eigenvalues, already allows to see 
whether a given fixed point is stable or unstable. 

In a second step we will obtain the 
exact results for the spectrum of eigenperturbations.

\subsection{First step: uniform approximation}

It is natural to define the total occupation probability
of specie $k$ at scale
$\Gamma= T \ln t$ as
$p^{\Gamma}_k = \int_{\eta_1,\eta_2} P^{\Gamma}_k(\eta_1,\eta_2)$.
The difficulty of the problem comes from the
fact that it does not satisfy a closed equation.
However, if one also introduces 
$p^{\Gamma}_k(0) = \int_{\eta_2} P^{\Gamma}_k(0,\eta_2)
=\int_{\eta_1} P^{\Gamma}_k(\eta_1,0)$, i.e
the occupation probability of specie $k$ of the valleys
just being decimated at $\Gamma$, one can obtain a
closed coupled equation by integration of (\ref{rgreadiff})
which reads:
\begin{eqnarray} \label{intereadiff}
\Gamma \partial_\Gamma p^{\Gamma}_k = 
2 ( - p^{\Gamma}_k(0) + W^k_{k_1,k_2} p^{\Gamma}_{k_1}(0) p^{\Gamma}_{k_2} )
\end{eqnarray}
It is then tempting to set, as an approximation, 
$p^{\Gamma}_k(0)=p^{\Gamma}_k$. This would be correct
at any of the fixed points (\ref{fixedpointeta}), but since we are studying 
convergence to a fixed point, it is an approximation
which amounts to neglect the dynamical correlations between
the deviations in the specie concentration and in the distribution
of barriers heigths. For this reason we call it the
''uniform approximation''. It yields the following
approximate closed RG equation for the $p^{\Gamma}_k$:
\begin{eqnarray}
\Gamma \partial_\Gamma p^{\Gamma}_k 
= 2 ( - p^{\Gamma}_k + W^k_{k_1,k_2} 
p^{\Gamma}_{k_1} p^{\Gamma}_{k_2})
\label{approximate}
\end{eqnarray}
which preserves the normalization condition $\sum p_k=1$.
This approximate flow has the same fixed points 
$p^{\Gamma}_k = p_k^*$ than
the true one (\ref{eq}). This equation, remarkably,
is reminiscent of a ''mean field type'' rate equation,
except that the role of ''time'' would be 
played by the variable $\ln ( T \ln t)$.

The relaxation of (\ref{approximate})
towards any of these fixed points is studied 
by setting $p_k^{\Gamma}=p_k^*+f_k^{\Gamma}$ and linearizing
for the small perturbation $f_k^{\Gamma}$
around the fixed point $p_k^*$. It yields, in terms
of the matrix M introduced
in (\ref{matrixM}):
\begin{eqnarray} \label{eqrelaxf}
\Gamma \partial_\Gamma f^{\Gamma}_k = 2 ( - f^{\Gamma}_k +
2 M_{k k'} f^{\Gamma}_{k'} )
\end{eqnarray}
and thus the convergence towards the fixed point has components behaving
as $\Gamma^{- \Lambda_{\alpha}}$ where the exponents are given in terms
of the eigenvalues $\mu_{\alpha}$ of the matrix M as
$\Lambda_\alpha = 2( 1 - 2 \mu_\alpha)$ with $\alpha=2,..n$. 

So this ``uniform approximation'' would
indicate that a fixed point is stable if
all $Re(\mu_\alpha) < 1/2$ for all $\alpha=2,..n$,
and unstable otherwise. Remarkably,
this stability criterion coincides with the exact result
{\it when the eigenvalues are real} as we will now show, 
even if the naive convergence 
eigenvalues $\Lambda_{\alpha}$ are not correct
(they are ``renormalized'' to larger
absolute values). 

\subsection{Second step: full dynamics near a fixed point}

Up to now we have studied the convergence of the
landscape alone (\ref{rgvalley}), and the convergence
of the $p^{\Gamma}$ within a uniform approximation
assuming $p^{\Gamma}(0)=p^{\Gamma}$.

We now study the full dynamics near a fixed point solution
of the full reaction-diffusion equations (\ref{rgreadiff}).
We will indeed find that there are some correlations
between deviations in total occupation probabilities
(from the fixed point concentrations) and deviations
in the barrier distribution profile
(from the simple fixed point shape $e^{-\eta}$), resulting
in deviations with respect to the uniform approximation.

We thus consider a perturbation around the fixed point of the form

\begin{eqnarray}
P^{\Gamma}_k(\eta_1,\eta_2) = ( p^*_k + c^\Gamma_k(\eta_1,\eta_2) )
e^{- (\eta_1 + \eta_2)}  
\end{eqnarray}
and linearize the equation for the perturbation
$c^{\Gamma}_k(\eta_1,\eta_2)$ 
\begin{eqnarray}
&& \Gamma \partial_\Gamma c^\Gamma_k (\eta_1,\eta_2)=
((1+\eta_1) \partial_{\eta_1} - \eta_1 + (1+\eta_2) \partial_{\eta_2} ) - \eta_2) 
c^\Gamma_k (\eta_1,\eta_2) \\
&& + M_{k_1,k_2} [ \int_0^{\eta_1} d \eta' (c^\Gamma_{k_1}(\eta',\eta_2)
 + c^\Gamma_{k_1}(\eta',0))
+ \int_0^{\eta_2} d\eta' (c^\Gamma_{k_1}(\eta_1,\eta') 
+ c^\Gamma_{k_1}(0,\eta') ]
\end{eqnarray}
where we have used the symmetry of the $W$, and the definition of
the matrix M (\ref{matrixM}).

Note that in the end we are interested into the behavior
of the species proportions 
\begin{eqnarray}  \label{epsilon}
p^{\Gamma}_k=\int_{\eta_1,\eta_2} P^\Gamma_k (\eta_1,\eta_2)
= p^*_k+\epsilon_k^{\Gamma} \qquad \hbox{where} \qquad \epsilon_k^{\Gamma} =
 \int_{\eta_1, \eta_2} c^\Gamma_k(\eta_1,\eta_2) e^{- (\eta_1 + \eta_2)}
\end{eqnarray}
The normalisation condition of course implies that 
$\sum_k \epsilon_k^{\Gamma}=0$.

 Decomposing $c^\Gamma_k(\eta_1,\eta_2)$
upon the eigenvectors $p_k^{\alpha}$ corresponding
to the eigenvalues $\mu_{\alpha}$ of the matrix M as 
$c^\Gamma_k(\eta_1,\eta_2)=\sum_{\alpha} c^\Gamma_\alpha(\eta_1,\eta_2)
p^{\alpha}_k$, we obtain decoupled equations for the 
coefficients $c^\Gamma_\alpha(\eta_1,\eta_2)$
\begin{eqnarray}  \label{relaxation}
&& \Gamma \partial_\Gamma c^\Gamma_\alpha(\eta_1,\eta_2) =
((1+\eta_1) \partial_{\eta_1} - \eta_1 + (1+\eta_2) \partial_{\eta_2} ) 
- \eta_2) c^\Gamma_\alpha(\eta_1,\eta_2)\\
&& \mu_\alpha [ \int_0^{\eta_1} d \eta' (c^\Gamma_{\alpha}(\eta',\eta_2)
 + c^\Gamma_{\alpha}(\eta',0))
+ \int_0^{\eta_2} d\eta' (c^\Gamma_{\alpha}(\eta_1,\eta') 
+ c^\Gamma_{\alpha}(0,\eta') ]
\end{eqnarray}

For a given $\mu_\alpha$, we look for solutions behaving as
$c^\Gamma_{\alpha} \sim \Gamma^{-\Phi_{\alpha}}$
and determine the exponent $\Phi_{\alpha}$ as a function of
the eigenvalue $\mu_{\alpha}$. Here, a priori both
$\mu_{\alpha}$ and $\Phi_{\alpha}$ can be complex.

Before we study this equation for general $\mu$ we will
first study the simpler cases $\mu=0$ and $\mu=1$. Note that
for associative processes this will be sufficient.

\bigskip

\subsubsection{Study for $\mu_\alpha=0$}

\medskip

This case is important for naively stable fixed
points of associative processes (which have all $\mu_\alpha=0$
for $\alpha=2,... N-1$).
The $\Gamma$ dependent equation (\ref{relaxation}) with $\mu_\alpha=0$ 
can be integrated out explicitly starting from its initial value
at $\Gamma_0$
\begin{eqnarray}
c_{\alpha}^{\Gamma}(\eta_1,\eta_2) = (\frac{\Gamma}{\Gamma_0})^2 
e^{- (\frac{\Gamma}{\Gamma_0}-1)(\eta_1+\eta_2+2)}
c_{\alpha}^{\Gamma_0}(\frac{\Gamma}{\Gamma_0}(1+\eta_1)-1,
\frac{\Gamma}{\Gamma_0}(1+\eta_2)-1)
\end{eqnarray}
and thus the convergence of proportions $p^{\Gamma}_k$
towards $p^*_k$ is governed by (\ref{epsilon})
\begin{eqnarray}
\epsilon_{\alpha}^{\Gamma} =
 \int_0^{\infty} d\eta_1 \int_0^{\infty} d\eta_2 
 c^\Gamma_\alpha(\eta_1,\eta_2) e^{- (\eta_1 + \eta_2) }
= \int_{\frac{\Gamma}{\Gamma_0}-1}^{\infty} dy_1 
\int_{\frac{\Gamma}{\Gamma_0}-1}^{\infty} dy_2
 c^{\Gamma_0}_\alpha (y_1,y_2) e^{- (y_1 + y_2) }
\label{exactmu0}
\end{eqnarray}
So for generic initial perturbations $c^{\Gamma_0}_\alpha (\eta_1,\eta_2)$
that are not exponentially growing as $\eta_{1,2} \to \infty$,
we obtain that the convergence towards the fixed point is
exponential in $\Gamma$.
Note that here this exact result is very different from
the naive approximation which would predict a convergence
as $\Gamma^{-\Lambda}$ with $\Lambda=2$. To understand
why this is so, one can compute from the exact solution
(\ref{exactmu0}) the ratio:
\begin{eqnarray}
\frac{p_k^{\Gamma}(0)-p_k^* } {p_k^{\Gamma}-p_k^*} \sim 
\frac{\Gamma} {\Gamma_0}
\end{eqnarray}
which is found to grow with $\Gamma$!
This is why the uniform approximation
is particularly bad for this case where it predicts a power-law
instead of the exponential convergence in $\Gamma$.

\bigskip

\subsubsection{study for $\mu_\alpha=1$}

\medskip

To study the dynamics (\ref{relaxation}) in the case $\mu_\alpha=1$, 
it is useful to introduce the function $h_{\alpha}^{\Gamma} (\eta_1,\eta_2) 
= \partial_{\eta_1} \partial_{\eta_2} c_{\alpha}^{\Gamma} (\eta_1,\eta_2)$,
since it satisfies the closed simpler equation
\begin{eqnarray}  \label{twoderiv}
&& \Gamma \partial_\Gamma h_{\alpha}^{\Gamma} (\eta_1,\eta_2)  =
((1+\eta_1) \partial_{\eta_1} + (1+\eta_2) \partial_{\eta_2} ) 
+ 2 - \eta_1 - \eta_2) h_{\alpha}^{\Gamma} (\eta_1,\eta_2) 
\end{eqnarray}
that gives after integration from an initial condition at $\Gamma_0$
\begin{eqnarray}  \label{hfinal}
h_{\alpha}^{\Gamma}(\eta_1,\eta_2) = 
 (\frac{\Gamma}{\Gamma_0})^4 
e^{- (\frac{\Gamma}{\Gamma_0}-1)(\eta_1+\eta_2+2)}
h_{\alpha}^{\Gamma_0}(\frac{\Gamma}{\Gamma_0}(1+\eta_1)-1,
\frac{\Gamma}{\Gamma_0}(1+\eta_2)-1)
\end{eqnarray}
and thus for initial conditions $h^{\Gamma_0}_\alpha (\eta_1,\eta_2)$
that are not exponentially growing at $\eta_{1,2} \to \infty$,
we obtain that the the function $h^{\Gamma}_\alpha$
converges towards 0 exponentially in $\Gamma$.
This means that the perturbations $c^{\Gamma}_\alpha$ 
converges exponentially in $\Gamma$ towards the decoupled
subspace 
\begin{eqnarray}
c^{\Gamma}_\alpha (\eta_1,\eta_2)=\psi^{\Gamma}_\alpha (\eta_1)
+\psi^{\Gamma}_\alpha (\eta_2)
\end{eqnarray}
We now study the convergence towards the fixed point
in that decoupled subspace to see if there are
solutions behaving as $\psi^{\Gamma}_\alpha (\eta) \sim \Gamma^{-\Phi_\alpha}
\psi_{\alpha} (\eta) $ : the equation for
$\psi_\alpha (\eta)$ reads

\begin{eqnarray} \label{eqpsi}
-\Phi_\alpha \psi_{\alpha} (\eta)
= [(1+\eta) \partial_{\eta} - \eta] \psi_{\alpha} (\eta)
+ 2 \int_0^{\eta} d \eta' \psi_{\alpha} (\eta') 
+ \eta \psi_{\alpha} (0)
\end{eqnarray}

The only well-behaved solutions are found to be:

\begin{eqnarray}
&& \Phi_\alpha = +1   \qquad \psi(\eta) = \psi_\alpha(0) ( 1 - \eta ) \\
&& \Phi_\alpha = -2   \qquad \psi(\eta) = \psi_\alpha(0) ( 1 + 2 \eta )
\label{2roots}
\end{eqnarray}

The first solution corresponds to the convergence as
$1/\Gamma$ of the landscape discussed
previously. This is not surprising since the linearized
RG equation (\ref{rgvalley}) is exactly (\ref{relaxation}) 
with $\mu=1$. Note en passant that (\ref{hfinal}) shows that small 
correlations between barriers in the same valley
decrease exponentially fast in $\Gamma$ towards the
subspace of statistically independent distributions.
Since this landscape eigenvector satisfies 
$\epsilon^{\Gamma}_{\alpha}=\int_{\eta_1,\eta_2} 
c^{\Gamma}_\alpha(\eta_1,
\eta_2) e^{-(\eta_1+\eta_2)}=0$
it does not affect the species proportions $p^\Gamma_k$.
It is the second eigenvector in (\ref{2roots})
which is relevant for the reaction diffusion processes
since $\epsilon^{\Gamma}_{\alpha} \neq 0$. It corresponds
to the unstable eigenvalue (growth as $\Gamma^2$) associated
to a naively unstable fixed point (e.g. of an associative process). 
Note that in that case, the unstable eigenvalue found $\Phi=-2$
coincides with the naive value $ \Lambda = 2 (1- 2 \mu)= -2$
of the uniform approximation. 

Physically, this eigenvalue can be understood for e.g. the process
(\ref{potts}). The unstable fixed point is the empty 
state E with $p_0=1$, $p_A=0$. Now if one starts 
at $t'$ very close to the fixed point,
there are very few $A$ and their number should not vary at first,
as they will rarely meet. This is indeed exactly what the above 
results says, namely:

\begin{eqnarray}
n_A(t) = \frac{1}{2} n_\Gamma p_A^\Gamma \sim \frac{1}{T^2 \ln^2 t} 
( p_A^*   +  (p_A(t')-p_A^*) \frac{\ln^2 t}{\ln^2 t'} )
\end{eqnarray}
and using that E has $p_A^* =0$.

\bigskip

\subsubsection{Study for general $\mu$}

\medskip

We now study the case of a general $\mu_\alpha$.
It turns out that one can find solution of the original equation 
(\ref{relaxation})
under the form 
\begin{eqnarray}
c^{\Gamma}_{\alpha}(\eta_1,\eta_2) = \Gamma^{-\Phi_{\alpha}} 
H_{\alpha}(\eta_1 + \eta_2)
\end{eqnarray}
where the function $H_{\alpha}(z)$ satisfies the differential equation

\begin{eqnarray}
0 = (2 + z) H''_{\alpha}(z) + (\Phi_{\alpha} + 1 - z) H'_{\alpha}(z) 
+ (2 \mu_{\alpha} -1) H_{\alpha}(z)
\end{eqnarray}
together with the boundary condition $2 H'_{\alpha}(0)
+\Phi_{\alpha} H_{\alpha}(0)=0$.

The only well-behaved solution at $z \to \infty$
 is the confluent hypergeometric function 
$H_\alpha(z) = U(1 - 2 \mu_\alpha , 3 + \Phi_\alpha, 2 + z)$,
and the boundary condition at $z=0$ determines the 
possible exponents $\Phi_{\alpha}$ that should satisfy :
$2 U'(1 - 2 \mu_\alpha , 3 + \Phi_\alpha, 2 )
+\Phi U(1 - 2 \mu_\alpha , 3 + \Phi_\alpha, 2 )
=0$. Using the identity
$z U'(A,B,z) - (z+1 - B) U(A,B,z) = - U(A-1,B-1,z)$,
this equation for $\Phi_{\alpha}$ reduces to

\begin{eqnarray}  \label{eqphi}
U(- 2 \mu_\alpha , 2 + \Phi_\alpha, 2 ) = 0
\end{eqnarray}

We now discuss the behaviour of the solutions
of this equation. One must distinguish two cases:

\subsubsection{real $\mu$}

\begin{figure}[tbh]

\centerline{\fig{10cm}{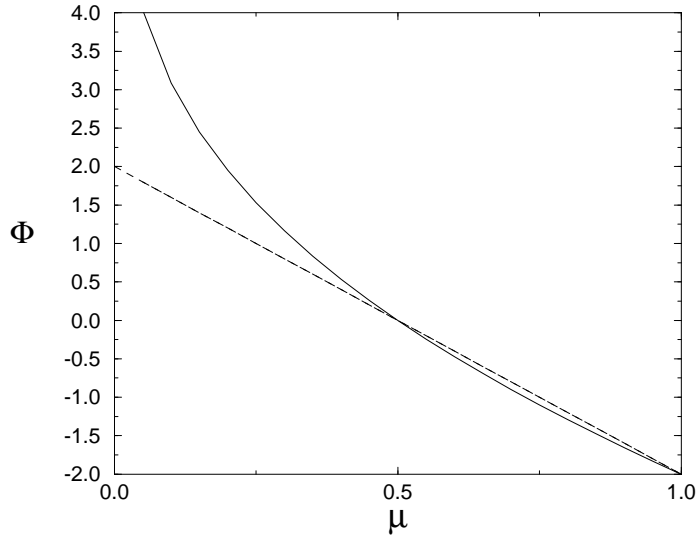}} 
\caption{\narrowtext  Plot of the exponent $\Phi$ 
($\Phi^{-}$) as a function of the eigenvalue $\mu$. It vanishes at
$\mu=1/2$ and diverges
as $\mu \to 0$. The result of the uniform approximation
is plotted as a dashed line. \label{figphi1}} 

\end{figure}

\begin{figure}[bth]

\centerline{\fig{10cm}{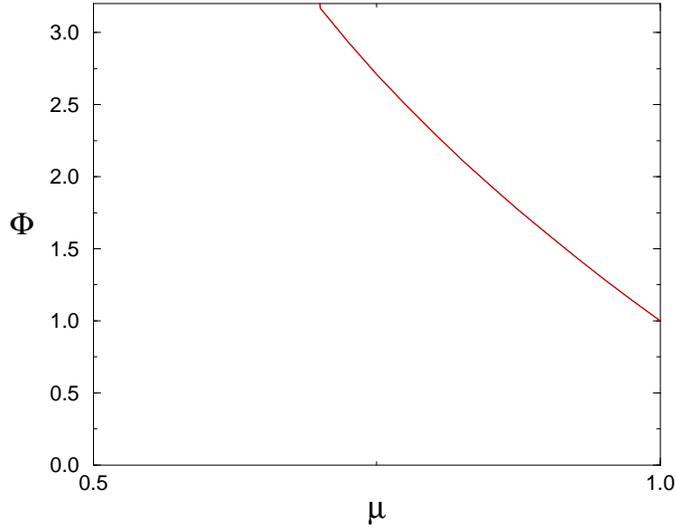}} 
\caption{\narrowtext  Plot of the exponent $\Phi^{+}$ 
as a function of the eigenvalue $\mu$. It diverges
at $\mu=1/2$ \label{figphi2}} 

\end{figure}

Let us start with $\mu$ real.
As mentionned above, one must have $\mu \leq 1$.
For $\mu=1$ the equation (\ref{eqphi}) reduces to
$\Phi^2 + \Phi -2=0$ which admits the two roots
$\Phi=1$ and $\Phi=-2$ and one thus recovers the eigenvalues
(\ref{2roots}). The equation (\ref{eqphi}) continues
to admit two finite roots when $\mu$ belongs to the interval
$\frac{1}{2} <\mu \leq 1$, which we denote 
$\Phi^{+}(\mu)$ and $\Phi^{-}(\mu)$ 
(with $\Phi^{+}(1)=1$ and $\Phi^{-}(1) = -2$).
The behaviour of these roots as a function of
$\mu$ is plotted in Fig. (\ref{figphi1}) and
(\ref{figphi2}). As $\mu$ is decreased from $1$,
$\Phi^{+}(1)$ increases and diverges when $\mu \to 1/2^{+}$
while $\Phi^{-}(\mu)$ increases from $-2$ to
$\Phi^{-}(1/2)=0$. For $\mu<1/2$ the equation (\ref{eqphi})
admits only one finite root $\Phi(\mu)=\Phi^{-}(\mu)$
which is positive and
with $\Phi(1/2)=0$ and $\Phi(\mu) \to +\infty$ 
as $\mu \to 0^{+}$.

Again we can see from the exact solution above, why the uniform 
approximation is not valid.
In terms of the dominant $\alpha$ mode
the ratio:
\begin{eqnarray}
\frac{p_k^{\Gamma}(0)-p_k^* } {p_k^{\Gamma}-p_k^*} \sim 
\frac{\int_0^{\infty} dz  H_{\alpha}(z) e^{-z} }
{\int_0^{\infty} dz  z  H_{\alpha}(z) e^{-z} } \equiv b_{\alpha}
\end{eqnarray}
goes at large $\Gamma$ towards 
the finite limit $b_{\alpha}$ found to be bigger than one
when $\mu < 1/2$.
One can check that setting $p_k^{\Gamma}=p_k^*+f_k^{\Gamma}$
and $p_k^{\Gamma}(0)=p_k^*+b f_k^{\Gamma}$ into
the equation (\ref{intereadiff}), one obtains
that the relaxation towards the fixed point is now
like $\Gamma^{- \lambda_{\alpha}(b)}$ with
$\lambda_\alpha (b)= 2( b - (b+1) \mu_\alpha)$.
Since $b_{\alpha}>1$, the correct exponent $\Phi_{\alpha}
=\lambda_\alpha (b_{\alpha})$ is bigger than
the naive exponent $\Lambda_{\alpha}
=\lambda_\alpha (b=1)$.

\subsubsection{complex $\mu$}

Next we turn to complex $\mu$. The matrix $M$ being real,
if it has complex eigenvalues they will come in pairs, $\mu_\alpha$ and
$\mu_\alpha^*$ corresponding to complex conjugate
eigenvectors $p^\alpha_k$ and $(p_k^\alpha)^*$.
In the uniform approximation they will combine 
as $f_k^\Gamma = \Gamma^{- Re(\Lambda_\alpha)} 
r_k \cos( Im(\Lambda_\alpha) \ln \Gamma + \phi^\alpha_k)$
and correspond to oscillatory (growing or decaying)
solutions (where $\Lambda = 2 (1 - 2 \mu)$ is complex).
This situation happens in cyclic reactions, examples of which
will be given below. Within the uniform approximation 
the fixed point is stable to this perturbation 
if $Re(\mu) < 1/2$ and unstable otherwise. This
however turns out {\it not} to be correct. Indeed the 
correct exponent $\Phi$ (now complex) is determined by
the above equation (\ref{eqphi}) associated to $\mu$. 
One has $\Phi(\mu^*) =\Phi(\mu)^*$,
and to each pair of $\mu$, $\mu^*$ one can associate
one (or two in some cases) pairs of $\Phi$, $\Phi^*$ 
also corresponding to oscillatory solutions. The oscillation
frequency in the $\ln \Gamma$ variable is now given by $Im(\Phi)$, 
and the stability being now determined by $Re(\Phi)$
($Re(\Phi) > 0$ corresponds to a stable eigenperturbation,
while $Re(\Phi) < 0$ correspond to an unstable one).

\begin{figure}[tbh]

\centerline{\fig{10cm}{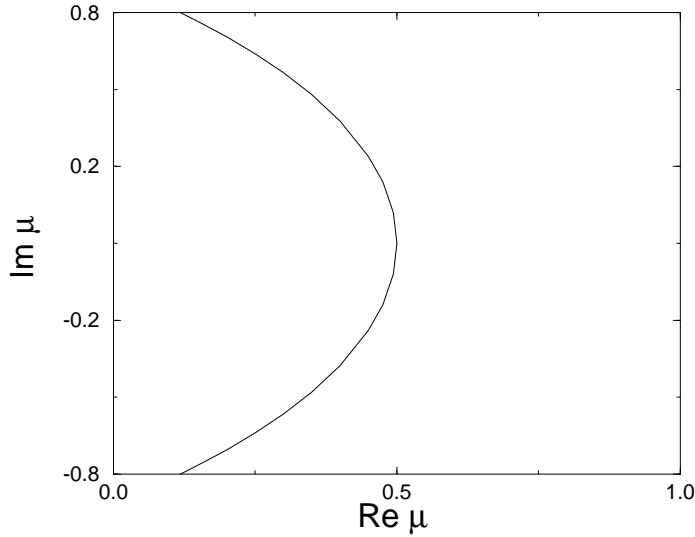}} 
\caption{\narrowtext Stability diagram: the solid line 
in the complex plane
of $\mu$ delimitates the region of instability
($Re(\Phi) < 0$ to the right of the line) from the region
of stability ( $Re(\Phi) > 0$ to the left) \label{figphistab}} 

\end{figure}

Interestingly, $Re(\Phi(\mu))$ is a decreasing function of
$Im(\mu)>0$ as it is increased from $0$. Thus the region of
stability in the complex $\mu$ plane 
is {\it different} from the one inferred
from the uniform approximation. It is represented in 
(\ref{figphistab}). One notes that for complex eigenvalues
one can have a fixed point naively stable (with $Re(\mu) < 1/2$) 
which is in reality unstable (with $Re(\Phi <0)$.

\bigskip

\subsection{asymptotic dynamics: conclusion }

Thus we have solved the problem of the dynamics near 
the asymptotic states of the form (\ref{eq})
for arbitrary reaction diffusion process. Let us
summarize the results.

\bigskip

{\it When the eigenvalues $\mu_{\alpha}$ of the matrix M (\ref{matrixM})
are real},
the stability of the fixed
point is determined by the naive argument :
a fixed point is stable if $\mu_\alpha <1/2$ for all $\alpha=2,..$.
The decay exponents $\Phi_\alpha$ are obtained in terms
of the $\mu_{\alpha}$ as :

\begin{eqnarray}
\Phi_\alpha  = F[\mu_\alpha]
\end{eqnarray}

where the function $F[x]$ is defined implicitly by the single root 
of the equation $U[-2 x , 2 + F[x], 2]=0$ (here $0<x<1/2)$
and represented in Fig. (\ref{figphi1}). 

In terms of these exponents, the large time behaviour for the concentrations
of the species is found to be
\begin{eqnarray} \label{concentrationPhi}
&& n_k(t) = \frac{1}{2} n_\Gamma p_k^\Gamma \\
&& p_k = p_k^* + \sum_{\alpha=2,N-1} \frac{b^\alpha_k}{(T \ln t)^{\Phi_\alpha}} + ...\\
&& n_\Gamma = \frac{2}{T^2 (\ln t)^2} ( 1 + \frac{c}{T (\ln t)} + ..)
\end{eqnarray}
where the $O(1/\Gamma)$ correction in $n_\Gamma$ comes from
the convergence of the landscape \cite{footnoteng}. In addition there are 
corrections to (\ref{concentrationPhi}) which decay must faster,
exponentially in $\Gamma$ (i.e algebraically in time).
The $b^\alpha_k$ are constants, depending on the initial condition.
The formula (\ref{concentrationPhi}) can also be used to relate two
late times. If the system is very near the asymptotic state at $t'$, with
$p_k(t')=p_k^* + \epsilon_k$, (\ref{concentrationPhi}) holds at $t$ with 
$b^\alpha_k = \epsilon_\alpha p^\alpha_k (T \ln t')^{-\Phi_\alpha}$
where we recall that the $p^\alpha_k$ are the eigenvectors of $M$ and 
$\epsilon_k = \sum_\alpha \epsilon_\alpha p^\alpha_k$.

For practical applications it is useful to note that
ratios of concentrations of different species 
involve only the exponents $\Phi_\alpha$. On the other hand,
because of the factor $n_\Gamma$,
the relaxation of the concentration of a single specie $k$
to its asymptotic form is controlled (provided $p_k^* >0$) 
by the exponent  
$\min(1, \Phi)$ (where $\Phi$ is the minimum of all
exponents $\Phi_\alpha$ appearing in the corresponding
formula (\ref{concentrationPhi}) for $n_k(t)$).
The formula is even more interesting in the case $p_k^*=0$
(i.e if the specie $k$ disappears in the reaction) since then
the first correction becomes the dominant decay and one has
at large time that $n_k(t) \sim 1/(T \ln t)^{2 + \Phi}$.
Examples of such cases are studied in Section \ref{coales}.

Let us stress again that the difference between the exact value $\Phi_\alpha$ and the
uniform approximation value $\Lambda_\alpha$ is due to the fact that 
near the asymptotic states the ratios $\frac{p^{\Gamma}_k(0)}{p^{\Gamma}_k}$
differ from one, i.e the valleys to be decimated do not
have the average distribution of species: there is a non trivial mixing
between valley heights and concentration of species, missed by the naive argument,
and responsible for the non trivial relaxation exponents found here.

\bigskip

{\it When some eigenvalues $\mu_{\alpha}$ of the matrix M (\ref{matrixM})
are complex} the fixed point is stable provided all $\mu_\alpha$ 
lie in the part of the complex plane on the left to the curve represented in
Fig. (\ref{figphistab}). The specie concentrations then relax
with oscillations as:

\begin{eqnarray} \label{concentrationc}
n_k(t) = \frac{1}{2} n_\Gamma 
( p_k^* + \sum_{\alpha=2,N-1} 
\frac{b^\alpha_k}{(T \ln t)^{Re(\Phi_\alpha)}} 
\cos( Im(\Phi_\alpha) \ln (T \ln t)
+ \phi^\alpha_k) + ...
\end{eqnarray}

Finally, in the case where $\mu=1/2$ (or more generally on the line
$Re(\Phi) =0$) linear analysis is insufficient to determine the
evolution of the system, and one must study the full non linear
RG equation (\ref{rgreadiff}), which goes beyond 
the present study \cite{footnotend}.

\bigskip

To conclude this Section, let us recall that the 
results derived above concern, strictly
speaking, the effective dynamics described by the 
RG equation (\ref{rgreadiff}). As was discussed
in great details in Ref. \cite{us_long} for the single particule
diffusion, there are corrections in the real dynamics, with respect
to the effective dynamics. Indeed, in the effective dynamics
the whole thermal packet jumps at $\ln t = \Gamma$
over a barrier $\Gamma$, while in the real dynamics
typically a fraction of a thermal packet (which can be written as
$1 - \exp(- e^{- \Gamma (1 - \frac{\ln t}{\Gamma})})$) 
has not yet jumped at time $t$.
Since the distribution of barriers becomes broader and
broader, this generates corrections which at large
time are only subdominant for most quantities
(at most $O(1/\Gamma)$) coming typically
from rare events such as degeneracy of order $O(T)$ 
between neighboring barriers. They become dominant only 
for certain quantities, such as the width of the thermal packet,
which have vanishing leading order in the effective dynamics.
In Ref. \cite{us_long} the corrections to first order in 
$O(1/\Gamma)$ were evaluated and found to originate from
three rare events (a) valleys with degenerate minima,
(b) almost degenerate barriers (c) valley just being decimated
with a barrier $\Gamma+\epsilon$ (see fig. 7 of \cite{us_long}).

A similar detailed study of the rare events in presence
of reaction processes can be performed but goes beyond the 
present paper. With similar arguments as in \cite{us_long}
we do not expect any correction to the leading order of the
quantities computed in this paper. In principle, subdominant
corrections could add to the subleading terms computed
above, e.g. in (\ref{concentrationPhi}). They are certainly at most
of order $O(1/\Gamma)$ (and thus cannot affect any decay
as $\Gamma^{- \Phi}$ with $\Phi <1$) but it is likely that
they are even of higher order. Indeed most of these corrections
(e.g. (b)) come from single particle diffusion and can be reabsorbed into
the global factor $n_\Gamma$. Other events (such as (a)) cannot affect
specie concentrations. Although this point deserve further study, it 
is likely that the corrections from the real dynamics to the convergence to 
asymptotic states obtained in this Section using is subdominant.

\medskip

\section{Examples of processes}

\label{examples}

Up to now we have only applied the general results only to
the process (\ref{potts}).
We give here several examples of other processes, 
starting with a process which exhibits a dynamical 
phase transition and to which the general results
apply directly. Then we present other cases which
raise interesting questions which go slightly beyond
our present analysis.

\subsection{a dynamical phase transition}

\begin{figure}[thb]

\centerline{\fig{10cm}{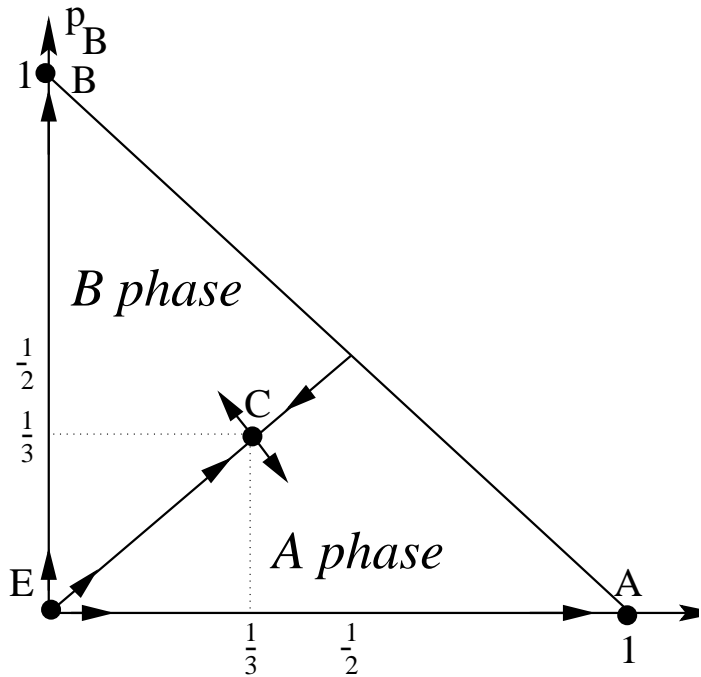}} 
\caption{\narrowtext 
Dynamical phase diagram of the reaction (\ref{reac1}) studied in the text.
\label{figreac1}} 

\end{figure}

\medskip

Let us consider the system involving two species A and B
and the empty state 0 :

\begin{eqnarray}
A+A \to A \qquad B+B \to B \qquad A+B \to 0
\label{reac1}
\end{eqnarray}

The solutions of equation (\ref{eq}) for the fixed points are 
the empty state $E=(p_A^*=0,p_B^*=0,p_0^*=1)$, the A-phase 
$(p_A^*=1,p_B^*=0,p_0^*=0)$, the B-phase $(p_A^*=0,p_B^*=1,p_0^*=0)$,
and the fixed point $C=(p_A^*=\frac{1}{3},p_B^*=\frac{1}{3},p_0^*=\frac{1}{3})$.

The A phase and the B phase are attractive fixed points with
eigenvalues $\mu_{2,3}=0$ corresponding to asymptotic exponential
decay in $\Gamma=T \ln t$ (i.e a power law in time)
of the other specie. The empty state
is a repulsive fixed point with eigenvalues $\mu_{1,2,3}=1$.
The critical point C is attractive for symmetric perturbation
$\delta p_A = \delta p_B= -\frac{\delta_0}{2}$ corresponding
to eigenvalue $\mu_3=0$,
but instable with eigenvalue $\mu_{2}=\frac{2}{3}$ for any dissymetric
perturbation $\delta p_A \neq \delta p_B$.
This corresponds to the exponent $\Phi^-(\frac{2}{3})=-0.761258$ 
(the other root is $\Phi^+(\frac{2}{3})=3.51853$). Since it is globally
attractive over the critical manifold, this fixed point controls the
dynamical transition from the $A$ phase to $B$ phase. Thus we conclude that
if one starts with a system of $A$ and $B$ in almost equal concentrations,
the difference $\vert p_A(t) -p_B(t) \vert$ (or equivalently the
relative concentrations of $A$ and $B$) grows with time as:

\begin{eqnarray}
\vert p_A(t) -p_B(t) \vert \sim (\ln t)^{\nu} \qquad \nu=0.761258
\label{reac1res}
\end{eqnarray}
or, equivalently, the differences of absolute concentrations decay
as $\vert N_A(t) -N_B(t) \vert \sim (\ln t)^{-2 + \nu}$,
i.e more slowly than the decay of both concentrations of
$A$ and $B$, which itself behaves as $(\ln t)^{-2}$.
The system eventually reaches a broken symmetry state where
either $A$ or $B$ predominates after a time $t_{br}$ which scales as:
\begin{eqnarray}
t_{br} \sim e^{c(t') \vert p_A(t') -p_B(t') \vert^{-1/\nu}}
\label{reac1res}
\end{eqnarray}
where $t'$ is a (shorter) reference time scale and $c(t')$ a ($t'$ dependent) constant.
Note that the uniform approximation would predict $\nu=2/3$ significantly
smaller that the exact result. Finally, the asymptotic final decay of the minority
phase is fast, exponential in $\Gamma$ (as $\mu=0$ at either $A$ ot $B$ fixed points).

Restraining from giving further examples among the large number of possible processes
with similarly interesting behaviour to which our general results readily apply,
we now turn instead to cases where open questions remain.

\medskip

\subsection{Reaction with a marginal fixed point}

\medskip

Let us now consider the reaction:

\begin{eqnarray}
A+A \to 0 \qquad B+B \to B \qquad A+B \to 0
\end{eqnarray}

The fixed points are the empty state $E(p_A^*=0,p_B^*=0,p_0^*=1)$
which is unstable with eigenvalues $\mu_{2,3}=1$,
the B phase $(p_A^*=0,p_B^*=1,p_0^*=0)$
which is fully attractive with eigenvalues $\mu_{2,3}=0$,
and the fixed point U $(p_A^*=\frac{1}{2},p_B^*=0,p_0^*=\frac{1}{2})$
which is attractive with $\mu_3=0$ along the
axis $p_B=0$ and marginal with $\mu_2=1/2$ for
any perturbation $ \delta p_B>0$.
Going beyond the linear approximation, we find
in the uniform approximation that $\Gamma \partial_{\Gamma} p_B^{\Gamma}
\simeq 2  (p_B^{\Gamma})^2$, i.e. a small initial proportion
$p_B^{\Gamma_0}$ grows very slowly as 
\begin{eqnarray}
 p_B^{\Gamma}
\simeq \frac{p_B^{\Gamma_0}}{1-2 p_B^{\Gamma_0} \ln (\frac{\Gamma}{\Gamma_0})}
\end{eqnarray}
and thus the time scale $t_{eq}$ where the proportion of $B$ becomes finite
grows like $\Gamma = T \ln t_{eq} \simeq e^{1/(2p_B^0)}$ for small
$p_B^{\Gamma_0}$. 
In the real dynamics, the exponent $\Phi(\mu=1/2)$ being
$0$ we also expect a kind of marginal behaviour near the fixed point
U. A full study of this behaviour is an interesting question
which goes beyond the present paper.

\subsection{cyclic reactions and complex eigenvalues}

Let us study the reaction:

\begin{eqnarray}
&& A+A \to A \qquad B+B \to B \qquad C+C \to C \\
&& A+B \to B \qquad B+C \to C \qquad C+A \to A
\end{eqnarray}
The solutions of (\ref{eq}) are the three pure phases
$p_A=1$, $p_B=1$, $p_C=1$ and the mixed state 
$p_A=p_B=p_C=\frac{1}{3}$. The pure phase $p_A=1$ is
stable ($\mu=0$) in the direction $\delta p_C=-\delta p_A>0$,
and unstable ($\mu=1$) in the direction $\delta p_B=-\delta p_A>0$.
The mixed fixed point $p_A=p_B=p_C=\frac{1}{3}$
has complex eigenvalues $\mu_{2,3}=(1\pm i/\sqrt{3})/2$,
leading to purely imaginary naive exponents $\Lambda_{2,3}=\pm i 2/\sqrt{3}$.
As can be seen on Fig \ref{figphistab},
the exact convergence exponents $\Phi_{2,3}$, solutions of (\ref{eqphi}),
have a negative real part, 
and thus the fixed point $p_A=p_B=p_C=\frac{1}{3}$ is also {\it unstable}.
This shows that the asymptotic behaviour of the system is more complex than
being described by a fixed point of type (\ref{eq}).

In fact, going back to the equations (\ref{approximate}) 
of the uniform approximation, and eliminating $p_C=1-p_A-p_B$,
we obtain that the flow equations for the two variables $(p_A,p_B)$
take the "divergence free" form
\begin{eqnarray}
&& \Gamma \partial_{\Gamma} p_A = 2 \partial_{p_B} f(p_A,p_B)  \\
&& \Gamma \partial_{\Gamma} p_A = - 2 \partial_{p_A} f(p_A,p_B)
\end{eqnarray}
with $f(p_A,p_B)= p_A p_B ( 1-p_A-p_B) $.
As a consequence, all starting points where the three
concentrations $(p_A,p_B,p_C)$ are non-zero, belong to 
closed flow lines $p_k^{uniform}(\Gamma)$ of constant value of $f(p_A,p_B)$.
Thus in the uniform approximation the asymptotic behaviour 
is always cyclic. 

This however does not carry to the real dynamics, beyond
the uniform approximation, since that one can check that 
these cycles $p_k^{uniform}(\Gamma) e^{-\eta_1-\eta_2}$ are {\it not} solutions 
of the RG equation (\ref{rgreadiff}). Thus the question of determining
the asymptotic behaviour of this problem is still open. A more complex
cyclic solution, or a new non trivial fixed point are among 
the possiblities.

We close this section by noting that one can also expect 
from \cite{nicolis_chaos,velikanov_chaotic,coullet_chaotic}
that reactions with a large enough number of species have chaotic
solutions at the level of the uniform approximation.
It would be interesting to investigate whether such 
chaotic solutions
could also exist in the RG and in the exact dynamics of
these disordered reaction diffusion problems.

\subsection{Reaction with an infinite number of states }

We now consider the much studied $A+B \to$ inert reaction, 
which, in the absence of disorder is known to
lead to segregation \cite{cheng_segregation} of the two species.

\begin{eqnarray}
A + A \to A + A \qquad B + B \to B + B \qquad A + B \to 0
\end{eqnarray}

We introduce the notations $A_0=0$, $A_m = m A$ and
$A_{-m} = m B$ ($m \ge 1$). The possible contents for a valley
are now the $A_m$ with $m \in Z$ and thus their number is infinite.
The reaction rules become with these notations

\begin{eqnarray}
A_k + A_p \to A_{k+p} 
\end{eqnarray}

So for the RG procedure, it is convenient to associate to each valley
an auxiliary variable $m$ representing the contents
of the valley, and to write the RG equation for
the probability distribution $P^{\Gamma}(z_1,z_2;m)$
where the RG rule for the auxiliary variable $m$ upon fusion of valleys
simply reads $m = m_1 + m_2$.
We can use the result of the Appendix of (\cite{danfisher_rg2}) (for the same
RG rule of an auxiliary variable) and obtain the scaling
\begin{eqnarray}
<m^2>  \sim \Gamma^2
\end{eqnarray}
Thus we find that charges of order $\Gamma=T \ln t$ of both signs 
(i.e groups of size of order $\Gamma=T \ln t$ of particles of the same specie)
will accumulate near the bottom (in a packet of typical size $O(1)$)
of each renormalized valley.
These packets will be separated by a large distance of order $(T \ln t)^2$.
The total number of particles will still decay, as $1/(T \ln t)$. 
This asymptotic state thus still presents strong features reminiscent of the
segregation observed in the pure case \cite{cheng_segregation}. By contrast 
with the pure model, here several packets of $A$ can also be found 
in successive neighboring valleys.

\section{persistence properties}

\label{persistence}

We now study persistence properties in the reaction 
diffusion models defined in (\ref{secmodel}).
As explained in the Introduction one is interested in
computing probabilities that some type of event has not occured
between time $0$ and $t$. The decay with time $t$ of these
probabilities usually involves new non trivial
exponents which characterize the non equilibrium dynamics.
Since they integrate over time the behaviour of the
system, they are usually hard to obtain analytically,
even in the pure systems. For reaction diffusion models
in a random environment the following types of persistence
probabilities can be defined, and will be studied in
the following corresponding sections.

A - The simplest persistence observable is the probability
$\Pi(t)$ over runs and environments that a given point 
$x=0$ has not been crossed by any particle between time $0$ and $t$
in a given run. The decay of $\Pi(t)$:
\begin{eqnarray}
\Pi(t) \sim \overline{l}(t)^{-\theta}
\label{refdef}
\end{eqnarray}
where $\overline{l}(t)$ is a characteristic length of the
diffusion process, here given from (\ref{scaling}) as 
$\overline{l}(t) = \frac{1}{2} (T \ln t)^2$, defines
the persistence exponent $\theta$.
This is the definition used in this paper, even when
referring to the pure case, where $\overline{l}(t) \sim
\sqrt{t}$, whereas another frequent definition is in
terms of the power of $t$. Since here the diffusion
is logarithmic we choose everywhere in this paper 
the more general definition 
(\ref{refdef}) both for pure and disordered problems.

B - In the presence of quenched disorder one can also 
study the probability $\Pi_{th}(t)$ over environments
that a given point $x=0$ has not been crossed by
any of the thermally averaged trajectories $<x(t)>$ 
of the particles. Similarly the decay of 
$\Pi_{th}(t) \sim \overline{l}(t)^{-\overline{\theta}}$
defines the exponent $\overline{\theta}$. One expects in general that
$\overline{\theta} \leq \theta$ and here we find that 
these two exponents are quite different.

C - More generally, one can defined the probability that a 
given pattern present at time $0$ has survived up to time 
$t$. We study the example of the survival of domains (i.e. intervals
between particles), which in the pure case was shown
 to lead to the definition of
two new exponents \cite{krapivsky_benaim} called $\delta$ and $\psi$ :
$\delta$ characterises the probability that a domain
has survived up to time $t$ without merging with other domains,
and $\psi$ characterises the probability that a domain
has survived up to time $t$ with mergings with other domains.

D - Finally we study the exponents $\delta_A$
and $\psi_A$ characterising the probability
that a particle A has survived up to time $t$,
without any coalescence and with coalescences respectively.

\subsection{Persistence in a single run}

\subsubsection{no crossing by any particles: exponent $\theta$}

We start by computing the probability
$\Pi(t)$ that $x=0$ has not been crossed by {\it any} particle
up to $t$. We consider a rather general reaction diffusion process
with a vacuum state $0$ ($k=0$) and occupied states (with particles
in them), $k \ge 1$. To solve this problem we can consider separately
the two half spaces $x>0$ and $x<0$ and study the problem of
a semi infinite system ($x>0$) with an absorbing boundary
at $x=0$ (absorbing for the states $k \ge 1$).

\begin{figure}[thb]

\centerline{\fig{12cm}{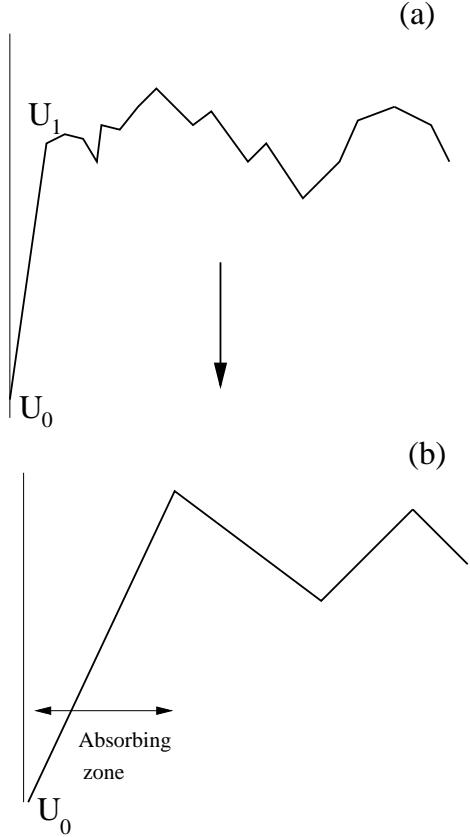}} 
\caption{\narrowtext 
Illustration of the RG in presence of an absorbing boundary.
(a) the boundary at site $x=0$ can be represented by setting $U_0=-\infty$.
(b) renormalized landscape, with the absorbing zone (see text).
\label{figboundary}} 

\end{figure}

For diffusion in a Sinai landscape in presence 
of an absorbing boundary at $x=0$ one defines a 
new RSRG with slightly new rules: the boundary RSRG,
explained in detail in Ref. \cite{us_long}.
The first bond is by definition always ascending
with an infinite barrier (and thus can never be decimated)
and represents an ``absorbing zone'' 
(see Figure (\ref{figboundary})). 
If the smallest barrier in the system at $\Gamma$
is the third bond from
the boundary or further the rules are identical to bulk RSRG.
If the smallest barrier is the second bond, i.e the first
descending bond, the procedure consists in eliminating the 
first valley (i.e the second and third bond) which is
merged with the absorbing zone.

Since the reaction rules of the species upon merging
valleys are unaffected by the boundary, at a given 
$\Gamma$ all renormalized valleys in the bulk
are distributed independently with 
$P^{\Gamma}_k(\eta_1,\eta_2)$ which satisfies the
equation (\ref{rgreadiff}). We now explicitly check
that the first renormalized valley has also the same
distribution. Indeed the probability $R_k(\eta,\eta')$ 
that the first renormalized valley has ($\eta$, $\eta'$, $k$)
satisfies the RG equation:
\begin{eqnarray} \label{rgfunctionR}
&& \left( \Gamma \partial_\Gamma - (1+ \eta) \partial_\eta 
- (1+ \eta') \partial_{\eta'} - 2 \right) R^{\Gamma}_k(\eta,\eta') \\
&& = P^{\Gamma}_k(\eta,\eta') \int d\eta_2 \sum_{k'} R^{\Gamma}_{k'}(0,\eta_2)
+
W^k_{k_1,k_2} 
R^{\Gamma}_{k_1}(.,0)*_{\eta} P^{\Gamma}_{k_2}(.,\eta') \\
&& + W^k_{k_1,k_2} 
R^{\Gamma}_{k_1}(\eta,.)*_{\eta'} P^{\Gamma}_{k_2}(0,.)
- R^{\Gamma}_k(\eta,\eta') 
\int d\eta_2 \sum_{k'} P^{\Gamma}_{k'}(0,\eta_2)
\end{eqnarray}
where the first term corresponds to the
decimation of the second bond (which results in the
increase of the absorbing zone) and the old second valley
becomes the new first renormalized valley. The third
and fourth terms correspond to the decimation of the
left bond of the second valley (the loss term must be
explicitly written on the l.h.s. to keep the distribution
$R$ correctly normalized to unity). One can now check that
$R^{\Gamma}_k(\eta,\eta') = P^{\Gamma}_k(\eta,\eta')$
is a consistent solution.

To compute the persistence exponent,
we now define $V_k(\eta,\eta')$ as the probability over all environments
that the boundary at $x=0$ has never been crossed by a walker
between time $0$ and $\Gamma$ {\it and} the first renormalized 
valley has ($\eta$, $\eta'$, $k$) at $\Gamma$. It satisfies
the RG equation:

\begin{eqnarray} \label{rgfunctionV}
&& \left( \Gamma \partial_\Gamma - (1+ \eta) \partial_\eta 
- (1+ \eta') \partial_{\eta'} - 2 \right) 
V^{\Gamma}_k(\eta,\eta') \\
&& = P^{\Gamma}_k(\eta,\eta') \int d\eta_2 V^{\Gamma}_0(0,\eta_2)
+  W^k_{k_1,k_2} 
V^{\Gamma}_{k_1}(.,0)*_{\eta} P^{\Gamma}_{k_2}(.,\eta') \\
&& + W^k_{k_1,k_2} 
V^{\Gamma}_{k_1}(\eta,.)*_{\eta'} P^{\Gamma}_{k_2}(0,.)
-V^{\Gamma}_k(\eta,\eta') 
\int d\eta_2 \sum_{k'} P^{\Gamma}_{k'}(0,\eta_2)
\end{eqnarray}
similar to (\ref{rgfunctionR}) except for the first term
which carries the restriction that the second bond can be decimated 
{\it only if} the first renormalized valley is empty
(since, if it contains a particle, this particle gets absorbed by the wall,
i.e crosses the origin). A consistent solution is simply

\begin{eqnarray}
V^{\Gamma}_k(\eta,\eta') = v_\Gamma P^{\Gamma}_k(\eta,\eta')
\end{eqnarray}
with 
\begin{eqnarray}
\Gamma \partial_\Gamma v_\Gamma = - v_\Gamma 
\int_{\eta_2} \left( \sum_{k'} P^{\Gamma}_{k'} (0,\eta_2) 
- P^{\Gamma}_{k=0}(0,\eta_2) \right)
 \end{eqnarray} 

We now use the fact that the system reaches for large $\Gamma$
an asymptotic state
corresponding to an attractive fixed point (\ref{fixedpointeta}),
and this leads to the asymptotic decay

\begin{eqnarray}
v_\Gamma \sim \Gamma^{- (1-p_0^*)}
\end{eqnarray}
Since the probability $\Pi(t)$ that the point $x=0$ has not
been crossed by any particle up to time $t$ on the
infinite line is the square of the corresponding probability for the
semi infinite problem we obtain:

\begin{eqnarray}
\Pi(t) \sim v_\Gamma^2 \sim \overline{l}(t)^{- \theta}
\end{eqnarray}
with $\overline{l}(t) = \frac{1}{2} (T \ln t)^2$ and the result
for the persistence exponent:
\begin{eqnarray} \label{restheta}
\theta = 1-p^*_0
\end{eqnarray}

As an example, we explicit the result for 
the process (\ref{potts}) where $p_0^*=\frac{r}{r+1}$:

\begin{eqnarray}
\theta_r = \frac{1}{1+r}
\end{eqnarray}
For $r=0$ where the particles $A$ always merge and occupy all valleys,
we recover the half-space exponent $\frac{1}{2} \theta(r=0) = \frac{1}{2}$
corresponding to the decay exponent of the probability of no return to
the origin for a single Sinai walker obtained in 
\cite{us_prl,us_long}.

\subsubsection{number of particles absorbed by a wall: 
generalized persistence}

A generalization of the persistence exponent $\theta$ 
can be defined for reaction-diffusion models 
on the semi-infinite line $x>0$ in the presence of an 
absorbing boundary at $x=0$. There one can compute the
probability $Q_{\Gamma}(n)$ that exactly $n$ particules
have been absorbed by the wall up to time $t$.
Generalising the approach of (\ref{rgfunctionV}),
we obtain that $Q_{\Gamma}(n)$ satisfies
at large $\Gamma$:

\begin{eqnarray}
&& \Gamma \partial_\Gamma Q_{\Gamma}(n) = (1-p_0^*) 
[ Q_{\Gamma}(n-1) - Q_{\Gamma}(n) ]  \qquad n \ge 1 \\
&& \Gamma \partial_\Gamma Q_{\Gamma}(0) = - (1-p_0^*) Q_{\Gamma}(0)
\end{eqnarray}

The RG equation for the generating function $Q_{\Gamma}(z)=\sum_n z^n Q_{\Gamma}(n)$ 

\begin{eqnarray}
\Gamma \partial_\Gamma Q_{\Gamma}(z) = - (1-p_0) (1-z) Q(z)
\end{eqnarray}
thus leads to the decay $Q_{\Gamma}(z) \sim \Gamma^{- (1-p_0) (1-z) }$.
Introducing the rescaled number of absorbed particules  

\begin{eqnarray}
g = \frac{n}{ \ln \Gamma}
\end{eqnarray}
and using as in \cite{us_prl,us_long} the saddle point
method, we find after a Legendre transform,
that the probability distribution
Prob(g) behaves as

\begin{eqnarray}
{ \rm Prob(g) } \sim \Gamma^{- 2 \omega(g) }
\end{eqnarray}
with the generalized persistence exponent:

\begin{eqnarray}
2 \omega(g) = (1-p^*_0) - g + g \ln \left( \frac{g}{1-p^*_0} \right)
\end{eqnarray}

For $g=0$, one recovers the persistence exponent of the
half space $\omega(0) = \theta/2$,
where $\theta$ is given by (\ref{restheta}).
$\omega(g)$ has a zero minimum at $g_a=(1-p_0)$ which is thus the
value that $g$ takes with probability one as $\Gamma \to \infty$

\begin{eqnarray}
\frac{n}{\ln (T \ln t)} = (1-p^*_0) \qquad \hbox{with probability one as }
\qquad t \to \infty
\end{eqnarray}

\bigskip

\subsection{persistence of thermally averaged trajectories}

\begin{figure}[thb]
\centerline{\fig{8cm}{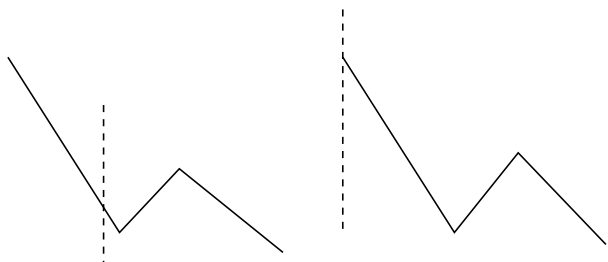}} 
\caption{\narrowtext If the point at $x=0$ 
(indicated by a dotted line) happens to lie in a renormalized 
valley as shown on the left figure,
it will be crossed many times by a single particle, while
it may not be crossed at all by the thermal average $<x(t)>$ 
before it jumps over the barrier on the right. This is not
the case in the situation shown on the right figure, where
typically no crossing of $x=0$ occurs for a single particle.
\label{fig3}}
\end{figure}

As was discussed in detail in Ref. \cite{us_long}, thermally averaged
trajectories of a single Sinai walker follow the effective dynamics
which we now use to study their persistence properties.
Fig. \ref{fig3} illustrates the difference in the persistence properties
between the single run dynamics studied in the previous Section and
the effective dynamics of thermally averaged trajectories.
Let us consider the case of a valley with a right bond of barrier $\Gamma$
such that the point $x=0$ lies to the left of a valley bottom
and is separated from it by a barrier less than $\Gamma$.
In this case, $x=0$ will be crossed (several times) in a typical single run,
while the thermal average $\langle x(t) \rangle$
will remain at the bottom of the valley until it jumps to the right
over the barrier without crossing $x=0$. Thus, as was found in
\cite{us_long} for the return to the origin of a single walker,
the exponents $\theta$ and $\overline{\theta}$ should generically
be different.

We now compute the probability $\Pi_{th}(t)$ that
the point $x=0$ has not been crossed by any particle
up to time $t$ within the effective dynamics. 
Let us define for each valley the auxiliary variables $m_1, m_2$
equal to the total number of sites in the descending ($m_1$)
and ascending ($m_2$) bonds respectively, which have not been 
crossed by any particle between $0$ and $t$. We define the
probabilities $P_k(\eta_1,\eta_2,m_1,m_2)$ that a valley has
a specie $k$, bonds $\eta_1$, $\eta_2$ and 
variables $m_1, m_2$. 
Consider the decimation represented in Fig. \ref{fig2}.
Let us denote the two valleys corresponding to bonds $(1,2)$ and $(3,4)$
containing respectively the species $k_1$ and $k_2$.
Upon decimation of bond $2$ the two valleys 
merge and the specie $k_1$ jumps to the bottom of the valley $(3,4)$
and thus goes over the bond $(2)$ and $(3)$ to react there with the
specie $k_2$. As a
consequence, the auxiliary variable of the new renormalized bond
$F_3'=F_1+F_3-\Gamma$ evolves with the rule:

\begin{eqnarray}  \label{rulempersistence}
m_3' = m_1 + \delta_{k_1,0} m_2 +\delta_{k_1,0} m_3 
\end{eqnarray}

This is a particular case of the auxiliary variables studied
in Appendix (\ref{auxiliary}) with $a_k=b_k=\delta_{k,0}$ and $d_k=1$.

The final result is that the fraction of sites that have never been
crossed by any particle in the effective dynamics decays as
$\frac{m}{\overline{l_{\Gamma}}}  \sim (\overline{l_{\Gamma}})^{-\overline{\theta}}$
where the persistence exponent $\overline{\theta}$ is solution of 
the following equation involving the confluent hypergeometric functions $U(a,b,z)$

\begin{eqnarray}
\overline{\theta}~~U(- p^*_0 , 2 \overline{\theta},1) 
= U(- p^*_0 , 2 \overline{\theta} + 1,1)
\label{eqthetabarre}
\end{eqnarray}

\begin{figure}[thb]
\centerline{\fig{12cm}{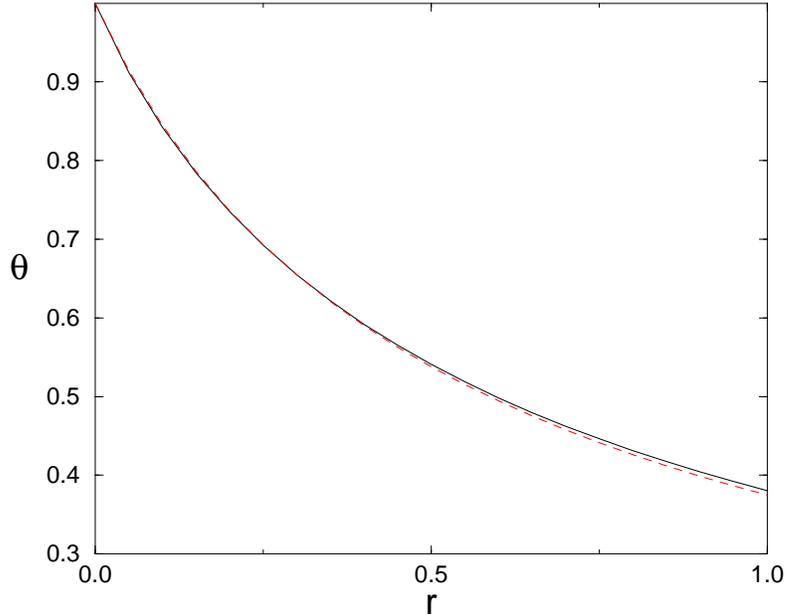}}
\caption{\narrowtext
Plot of $\overline{\theta}(r)$ as a function of $r$ (solid line)
and, for comparison, $\frac{1}{2} \theta_{pure}(r)$ (dashed line)
\label{fig22}}
\end{figure}

\begin{figure}[thb]
\centerline{\fig{12cm}{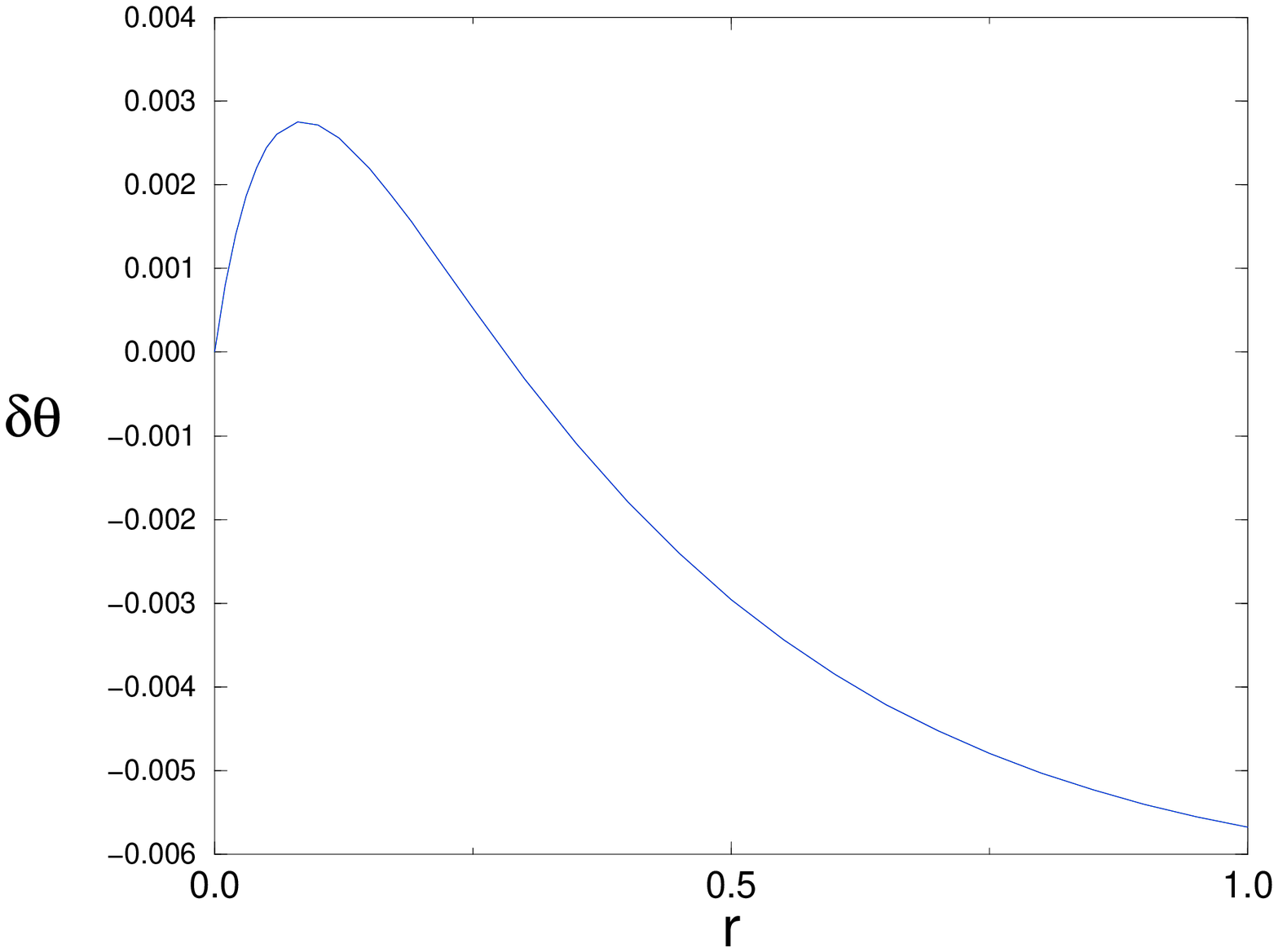}}
\caption{\narrowtext
Plot of $\delta \theta(r) = \frac{1}{2} \theta_{pure}(r) - \overline{\theta}(r)$:
the difference remains very small for all $r$. It
vanishes at $r=0$ and $r=0.280701..$ \label{fig23}}
\end{figure}

For the process (\ref{potts}) one has $p^*_0=r/(1+r)$,
and the resulting exponent $\overline{\theta}(r)$ 
is plotted in Fig. \ref{fig22}. Surprisingly
we find that it is numerically extremely close
for all $r$ (to less than about one percent in relative 
difference) of {\it one half} the result \cite{derrida_hakim} for
the pure system which reads
$\frac{1}{2} \theta_{pure}(r)
=-\frac{1}{8} + \frac{2}{\pi^2} (\arccos(\frac{r-1}{\sqrt{2} 
(r+1)}))^2$ and is also plotted in Fig. \ref{fig22}.
The expansion for small $r$ gives
\begin{eqnarray}
&& \overline{\theta}(r) = 1-2 r +o(r)  \label{eqthetabarre2} \\
&& \frac{1}{2} \theta_{pure}(r) =1- \frac{6}{\pi}  r +o(r)
\end{eqnarray}
and thus they are definitively different.
In the case $r=1$ where particles always annihilate, we obtain
\begin{eqnarray}
 \overline{\theta}(r=1) =0.380678..
\end{eqnarray}
which may be compared with $\frac{1}{2} \theta_{pure}(r=1)=3/8=0.375$.
The difference remains very small for all $r$, as shown in Fig. 
\ref{fig23}.

We have also generalized the calculation presented in this Section to 
compute the {\it number of visits} of thermally averaged trajectories of particles
at a given point. It leads again to a multifractal spectrum of exponents.
The calculation and the results are presented in the Appendix \ref{auxiliary}.

\subsection{Statistics of merging domains }

\begin{figure}[thb]
\centerline{\fig{8cm}{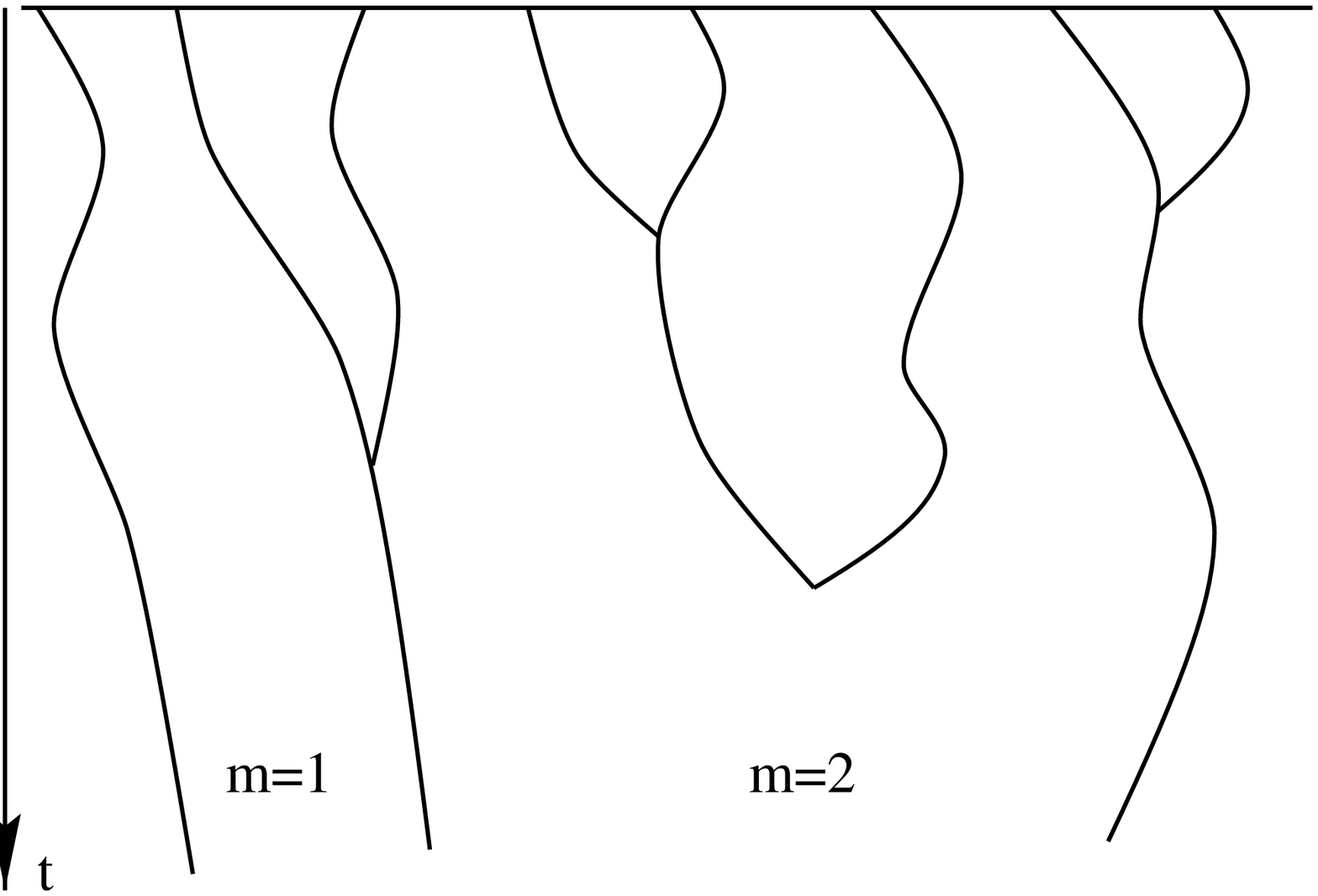}}
\caption{\narrowtext diffusion and merging of domains. To each domain
present at time $t$ (bottom of the figure) is associated the number
$m$ of ancestor domains in the initial state (defined at the top slice $t=0$).
Note that some initial domains die without producing any descendant domain at time $t$.
\label{figdomains}}
\end{figure}

Here we define domains as intervals between particles.
For the reaction (\ref{potts}),
on which we now concentrate, when a domain
dies (the two particules A meet)
the two contiguous domains can either merge if the particles annihilate
(with probability $r$) or remain separate if the particles coalesce
(with probability $1-r$). To characterize the statictics of 
the coarsening of domains in the pure case (i.e Potts domains with $q=1+\frac{1}{r}$)
Krapivsky and Ben-Naim have introduced \cite{krapivsky_benaim} 
the following definition. They define $Q_m(t)$ as the number of domains 
at time $t$ which have for ancestors $m \ge 1$ initial domains. 
This is illustrated on the figure \ref{figdomains}.

The total number $N(t)$ of domains remaining at time $t$ (equal to the total number of $A$
at time $t$) is simply given by the sum $N(t)=\sum_{m=1}^{+\infty} Q_m(t)$
and decays as $\sim p_A^*/\overline{l}(t)$, where $\overline{l}(t) \sim (T \ln t)^2$ 
is the characteristic length at time $t$, given in (\ref{scaling}).
The fraction of initial domains which have a descendant
that is still alive at $t$ is given by $S(t) = \sum_m m Q_m(t) = <m> N(t)$.
The decay of these quantities define two new independent persistence exponents
$\delta$ and $\psi$:
\begin{eqnarray}
&& Q_1(t) \sim \overline{l}(t)^{- \delta} \\
&& S(t) \sim \overline{l}(t)^{- \psi} 
\end{eqnarray}
and with these exponents, it is expected that
$Q_m(t)$ takes the scaling form 
\begin{eqnarray}
Q_m(t) = \frac{1}{\overline{l}(t)^{2-\psi} }{\cal Q}
\left(\frac{m}{\overline{l}(t)^{1-\psi}}\right)
\end{eqnarray}
The scaling function is expected to behave for small $z$ as
${\cal Q}(z) \sim z^{\sigma}$ where the exponent $\sigma$
is related to $(\delta,\psi)$ by the relation 
$\delta=2 - \psi+ (1-\psi) \sigma$. Note that the inequalities
$Q_1(t) \le \sum_{m} Q_m(t) \le \sum_{m} m Q_m(t)$ 
imply that $\psi \le 1 \le \delta$. Note that here, we 
have again defined the exponents with respect to the characteristic length
$\overline{l}(t)$ at time $t$. Thus in the pure case our definition
differs from the one of \cite{krapivsky_benaim} by a factor of $2$.

Below, we obtain the exponents $\delta(r)$ and $\psi(r)$
exactly for the process (\ref{potts}).

\subsubsection{Exponent $\delta(r)$ for the process (\ref{potts})}

For a domain to survive up to time $t$ while keeping its
variable $m=1$, the two domains walls must avoid meeting each other
up to time t, but they can meet other exterior domain walls, 
provided that upon meeting they coalesce and do not annihilate.

Since the two domains walls must not meet, given the properties
of the effective dynamics in the RG, the decay of $Q_1(t)$
is governed by the events such that at some $\Gamma_0$
the two domains belong to two neighboring renormalized valleys.
At all later times they will still belong to two neighboring renormalized valleys
and no decimation of the two renormalized bonds separating the
two domains can occur. As a consequence to compute the exponent $\delta$, 
we can consider separately the two corresponding half spaces.

For a given half space, we introduce the probability
$R^{\Gamma}(\eta,\eta')$ 
that{\it the first bond has never been decimated} and 
{\it the valley is ($\eta$, $\eta'$)}
and {\it there is always one walker in the first valley}.
The RG equation for this quantity reads
\begin{eqnarray}
&& ( \Gamma \partial_\Gamma - (1+ \eta) \partial_\eta 
- (1+ \eta') \partial_{\eta'} - 2 ) R^{\Gamma}(\eta,\eta') \\
&& = 
R^{\Gamma}(\eta,.)*_{\eta'} ( P^{\Gamma}_{0}(0,.)
 + (1-r) P_A^{\Gamma}(0,.) ) \\
&&  + R^{\Gamma}(.,0)*_{\eta} ( P^{\Gamma}_{0}(.,\eta'))
 + (1-r) P_A^{\Gamma}(.,\eta')) ) \\
&& - R^{\Gamma}(\eta,\eta') \int_0^{\infty} d\eta_2 \sum_{k'} P_{k'}^{\Gamma}(0,\eta_2)
\end{eqnarray}
where the $-$ term arises because the $R^{\Gamma}$, unlike the $P_k^{\Gamma}$
is not normalized to 1, and one must count the loss associated
with the left bond of the second valley. 
Integrating over $(\eta,\eta')$ one finds that 
$R^{\Gamma}=\int d\eta d\eta' R^{\Gamma}(\eta ,\eta')$ evolves with

\begin{eqnarray}
 \Gamma \frac{d R^{\Gamma}}{d\Gamma}  = - \int d\eta' R^{\Gamma}(0,\eta') 
- r  p_A^{\Gamma} \int d\eta R^{\Gamma}(\eta,0)
 - r   R^{\Gamma} \int_0^{\infty} d\eta  P_{A}^{\Gamma}(0,\eta)
\end{eqnarray}
corresponding to the three forbidden cases: decimation of the first bond, 
decimation of the second or third bond when both valleys 
are full and annihilation occurs.

The exponent $\delta$ will be given by the decay of the
half space probability $R(\eta,\eta')  \sim  \Gamma^{-\delta} $, since
the probability associated the two sides will decay as the square of the
probility for one side, i.e as $\Gamma^{-2 \delta}
\sim \overline{l}_{\Gamma}^{-\delta} $. Setting $R(\eta,\eta')  = \Gamma^{-\delta} 
e^{-\eta-\eta'} \rho(\eta,\eta')$, and using 
$p_0^* + (1-r) p_A^* =\frac{1}{1+r}$, one finds:

\begin{eqnarray}
&& 0= \left[ (1+\eta) \partial_\eta  + (1+\eta') \partial_{\eta'} 
+ (\delta-1-\eta-\eta') \right] \rho(\eta,\eta') + \\
&& \frac{1}{1+r} ( \int_0^\eta d\eta_1 \rho(\eta_1,0)
+ \int_0^{\eta'} d\eta_2 \rho(\eta,\eta_2) )
\end{eqnarray}

\begin{figure}[thb]
\centerline{\fig{12cm}{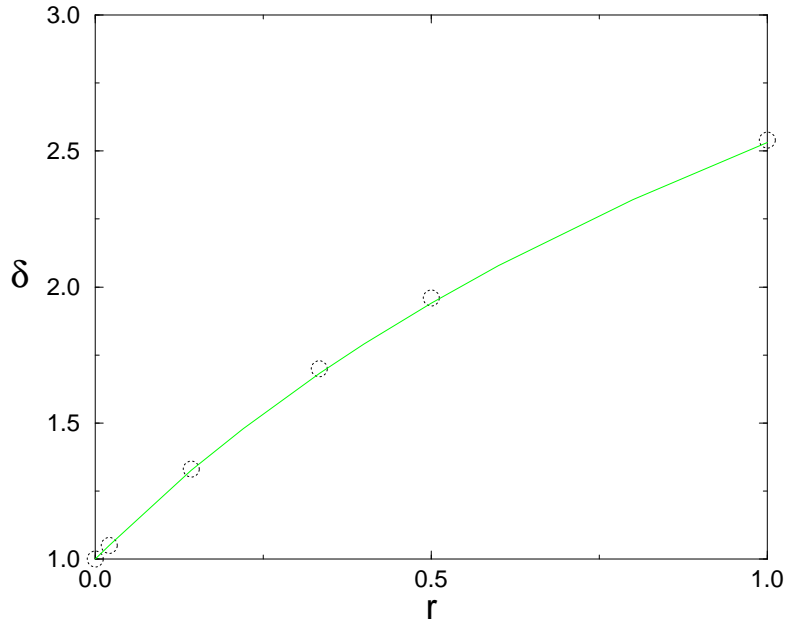}}
\caption{\narrowtext Plot of the exponent $\delta(r)$ for the process
(\ref{potts}) in the Sinai landscape, determined by equation
(\ref{deltaeq}) (solid line), and comparison with the
numerical results of Krapivsky and Ben Naim (Ref. 46) for the
pure case (circles). \label{figdeltar} }
\end{figure}

There is a solution of the form $\rho(\eta,\eta')=\tilde{\rho}(\eta + \eta')$
where $\tilde{\rho}(\eta)$ satisfies:

 \begin{eqnarray}
 0= \left[ (2+\eta) \partial_\eta  
+ (\delta-1-\eta) \right] \tilde{\rho} (\eta) +
 \frac{1}{1+r}  \int_0^\eta d\eta' \tilde{\rho}(\eta) 
\end{eqnarray}
After derivation with respect to $\eta$, one finds a standard
hypergeometric differential equation which admits for 
only solution not growing exponentially at $\eta \to \infty$,
the confluent hypergeometric function $U(\frac{r}{r+1},2+\delta,2+\eta)$.
The boundary condition at $\eta=0$ then determines the exponent $\delta(r)$
as the solution of the implicit equation:
\begin{eqnarray}
2 U'(\frac{r}{r+1},2+\delta(r),2)+(\delta(r)-1) 
U(\frac{r}{r+1}, 2 + \delta(r),2 ) =0
\end{eqnarray}
Using the functional relation 
$z U'(A,B,z)+(B-1-z) U(A,B,z)=-U(A-1,B-1,z) $, 
the exponent $\delta(r)$ is finally the solution of the equation
\begin{eqnarray}
 U(- \frac{1}{1+r}, 1+ \delta(r),2) =0
\label{deltaeq}
\end{eqnarray}
The solution of this equation is plotted in Fig. \ref{figdeltar}.

In the case $r=0$, we find $\delta(r=0)=1$ as expected.
Indeed, in that case where particles always coalesce, 
domains cannot merge, and thus $m=1$
is the only possible value : $Q_m(t)=\frac{1}{\overline{l} } \delta_{m,1}$
and thus $\delta=1=\psi$ as in the pure case.
For the case $r=1$, where particles always annihilate, we find 
\begin{eqnarray}
\delta(r=1) = 2.53083..
\end{eqnarray}
which is remarkably close to the numerical result obtained in
(\cite{krapivsky_benaim}) for the Ising pure case :
$\delta_{pure}(r=1) = 2.54(4)$. This puzzling feature
also holds for other values of $r$, 
as shown in Fig \ref{figdeltar}, with less than about one percent in
relative difference.

\bigskip

\subsubsection{Exponent $\psi(r)$ for the process (\ref{potts})}

To compute $\psi$ we need to obtain the scaling behaviour
of the average number of ancestors of the
domain $<m> \sim \Gamma^{2(1-\psi)}$. However this is a priori difficult,
as the variable $m$ is associated to a domain which can extend over many
renormalized bonds and is thus ``non local''.  However we can circumvent
this difficulty
by decomposing $m$ upon the several renormalized bonds
which make up a domain, in order to have a {\it local} rule under RG
for an auxiliary variable $b$ associated to bonds. Thus we write each variable $m$ for a domain made out of $q$ bonds, as the sum $m=b_1+b_2+..+b_q$, of new auxiliary variables,
each associated to a bond. Since $q$ does not grow with $\Gamma$, the
scaling of $<m>$ and $<b>$ with $\Gamma$ are identical.
We define RG rules for the local $b$ variables as follows.
We consider two neighboring valleys as in Fig \ref{fig2} with bonds
(1,2) containing specie $k_1$ and (3,4) containing specie $k_2$, 
respectively. One must think of the variable $b$ as counting the number of
ancestors associated to a renormalized bond and thus
upon decimation of bond (2) the variables $b$, $b'$ associated
to the new bonds of barriers $F_1 + F_3 - \Gamma$ and $F_4$ 
become:

\begin{eqnarray}
 b = b_1 + b_2 + b_3 ~~~\hbox{and}~~~ b'=b_4 ~~~ &\hbox{if}& ~~~~ k_1=\emptyset \\
 b = b_1  ~~~\hbox{and}~~~ b'=b_2 + b_3 + b_4 ~~~ &\hbox{if}&
~~ k_1=A ~~ \hbox{and}~~ k_2=\emptyset \\
 b = b_1  ~~~ \hbox{and}~~~ b'=b_4 ~~~~~~~~~ &\hbox{if}& ~~ k_1=A ~~ \hbox{and} ~~ k_2=A
\end{eqnarray}
The first case where the decimated valley is empty is obvious.
In the second case, where a particle $A$ jumps from valley (1,2)
to the empty valley (3,4), the ancestors of the domain to the right
of $A$ previously associated to the bonds (2), (3) and (4) must now all be
associated to the bond (4). In the third case, where the two $A$ meet,
the ancestors of (2,3) disappear of the problem in all cases (i.e annihilation
or coalescence of the $A$ particles).

Introducing the rescaled variables $\beta=b/\Gamma^\Phi$, where $\Phi=2(1-\psi)$, 
the fixed point RG equation for
the valley distribution $P_k(\eta,\eta',\beta,\beta')$ 
reads:

\begin{eqnarray}
&& 0 = ( (1+ \eta) \partial_{\eta} + (1+ \eta') \partial_{\eta'} 
+ 2 + \phi (\beta \partial_{\beta} + \beta' \partial_{\beta'} + 2) )
P_k(\eta,\eta',\beta,\beta') \\
&& + 
W^k_{k_1,k_2}
\int_{\beta_i} P_{k_1}(.,0,\beta_1,\beta_2)*_{\eta}P_{k_2}(.,\eta',\beta_3,\beta_4) \\
&& \delta(\beta-(\beta_1 + (\beta_2+ \beta_3) \delta_{k_1,\emptyset}))
\delta(\beta'-(\beta_4 + (\beta_2+ \beta_3) \delta_{k_1,A} \delta_{k_2,\emptyset})) \\
&&  + 
W^k_{k_1,k_2}
\int_{\beta_i} P_{k_1}(\eta,.,\beta_1,\beta_2)*_{\eta'}P_{k_2}(0,.,\beta_3,\beta_4) \\
&& \delta(\beta-(\beta_1 + (\beta_2+ \beta_3) \delta_{k_1,\emptyset} \delta_{k_2,A}))
\delta(\beta'-(\beta_4 + (\beta_2+ \beta_3) \delta_{k_2,\emptyset})) 
\end{eqnarray}

We introduce the two first moments ($k=0$ and $k=A$)
\begin{eqnarray}
 g_k(\eta,\eta') = \frac{1}{P^*(\eta,\eta')} \int d\beta d\beta' \beta P_k(\eta,\eta',\beta,\beta') 
\end{eqnarray}
where $P^*(\eta,\eta')=e^{-\eta-\eta'}$.

Using the symmetry $P_k(\eta,\eta',\beta,\beta')=P_k(\eta',\eta,\beta',\beta)$,
we find that they satisfy the closed system:

\begin{eqnarray}
&& 0 = 
( (1+ \eta) \partial_{\eta} + (1+ \eta') \partial_{\eta'} 
- \eta - \eta' - \Phi) g_0(\eta,\eta') \\
&&
+ p_0^* \int_0^\eta [ g_0(.,\eta') +g_0(.,0)+ g_0(0,.) +  g_A(.,0) ]
+  p_0^* \int_0^{\eta'} [g_0(\eta,.) + g_A(\eta,.)] \\
&& 0 = 
( (1+ \eta) \partial_{\eta} + (1+ \eta') \partial_{\eta'} 
- \eta - \eta' - \Phi) g_A(\eta,\eta') \\
&&
+ p_A^* \int_0^\eta [g_A(.,0) + g_0(.,0) +g_0(0,.)] 
+p_0^* \int_0^\eta  g_A(.,\eta') \\
&&
+ p_A^* \int_0^{\eta'} [g_A(\eta,.)+g_0(\eta,.)+g_0(.,\eta)]
+p_0^* \int_0^{\eta'}  g_A(0,.) 
\label{systemg}
\end{eqnarray}
The exponent $\Phi$ is determined by the condition that 
the solutions $g_0(\eta,\eta')$ and $g_A(\eta,\eta')$
of this system should be well behaved at infinity
(i.e. should not be exponentially growing).

We found that setting
\begin{eqnarray}
&& g_A(\eta,\eta')=S_A(z=\eta+\eta') \\
&& g_0(\eta,\eta')+g_0(\eta',\eta)=S_0(z=\eta+\eta') 
\end{eqnarray}
allows to obtain the following closed system
for the two functions $S_0(z)$ and $S_A(z)$ 
\begin{eqnarray}
&& 0=(2+z) S_0'(z)-(z+\Phi) S_0(z) +2 p_0^* \int_0^z [S_0(.)+S_A(z)] \\
&& 0=(2+z) S_A'(z)-(z+\Phi) S_A(z) + \int_0^z [S_A(.)+p_A^* S_0(z)]
\end{eqnarray}
i.e, independently of the antisymmetric part of $g_0(\eta,\eta')$ 
which we will not need. To decouple this system, we introduce 
two linear combinations $S_{\pm}(z)=c_A S_A(z)+
c_0 S_0(z)$ that satisfy closed equations
\begin{eqnarray}
0= (2+z) S_{\pm}'(z)-(z+\Phi) S_{\pm}(z) + \nu_{\pm}(r) \int_0^z S_{\pm}(.)
\label{systpsi}
\end{eqnarray}
where the eigenvalues, using $p_0^*=r/(1+r)=1-p_A^*$, are:
\begin{eqnarray}
\nu_{\pm}(r)=\frac{1}{2}+\frac{r}{1+r} \pm \frac{\sqrt{1+6r+r^2}}{2 (r+1)}
\end{eqnarray}
The only solutions of (\ref{systpsi})
that are not exponentially growing at infinity
are again given in terms of degenerate hypergeometric function
\begin{eqnarray}
S_{\pm}(z) \propto U(1-\nu_{\pm},3-\Phi,2+z)
\end{eqnarray}
The boundary conditions $2 S_{\pm}'(0)=\Phi S_{\pm}(0)$ finally give
\begin{eqnarray}
 U(-\nu_{\pm}(r),2\psi_{\pm}(r),2) = 0
\end{eqnarray}
 Since $\psi_+(r) < \psi_-(r)$,
the growth of $<m> \propto \Gamma^{2(1-\psi)}$
will be governed by $\psi_+(r)$, and thus the final result
is that the exponent $\psi(r)$ is determined by the equation
 \begin{eqnarray}
 U(-\frac{1}{2}-\frac{r}{1+r} -\frac{\sqrt{1+6r+r^2}}{2 (r+1)},2\psi(r),2)
= 0
\label{psieq}
\end{eqnarray}

In particular we have the following expansion around $r=0$
\begin{eqnarray}
\psi(r)=  1-\frac{5}{2} r +o(r)
 \end{eqnarray}

\begin{figure}[thb]
\centerline{\fig{12cm}{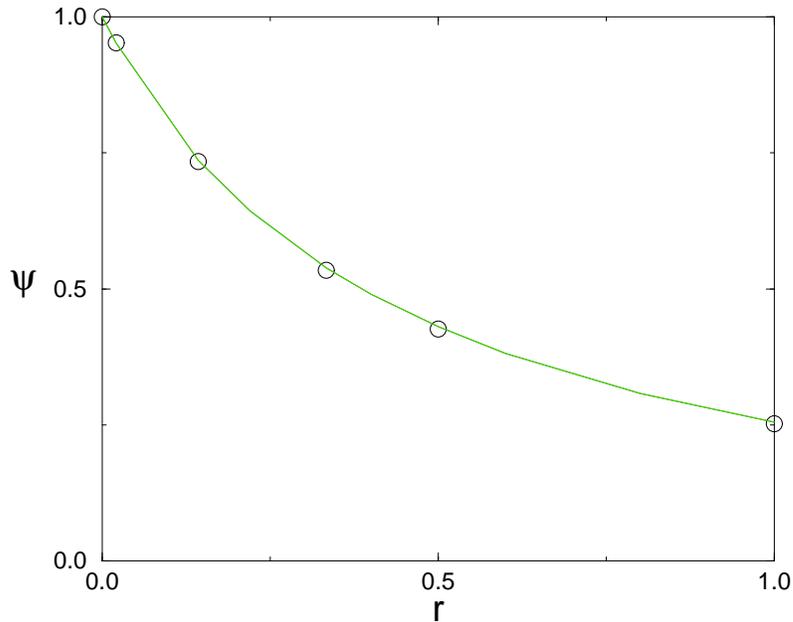}}
\caption{\narrowtext
Plot of the exponent $\psi(r)$ for the process
(\ref{potts}) in the Sinai landscape, determined by equation
(\ref{psieq}) (solid line), and comparison with the
numerical results of Krapivsky and Ben Naim (Ref. 46) for the
pure case (circles).\label{figpsir} }
\end{figure}

For the case $r=1$ where particles always annihilate,
we find
\begin{eqnarray}
\psi(r=1) = 0.254821..
\end{eqnarray}
which again is remarkably close to the numerical result obtained in
\cite{krapivsky_benaim} for the Ising pure case:
$\psi_{pure}(r=1) = 0.252(2)$, a property which hold
again for all $r$, and again to within less than about one percent
in relative difference, as illustrated in Fig \ref{figpsir}.

\begin{figure}[thb]
\centerline{\fig{12cm}{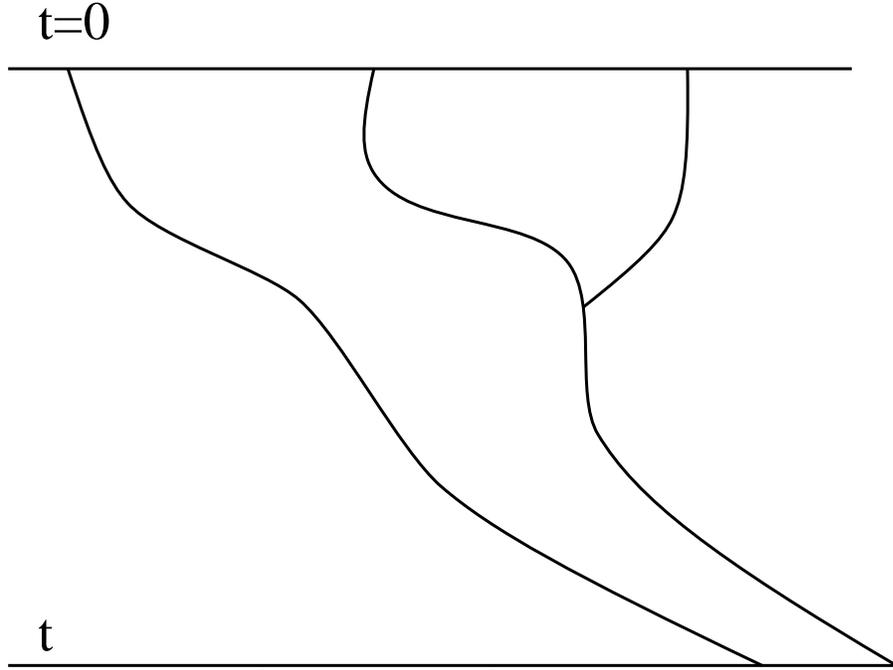}}
\caption{\narrowtext configuration where a surviving domain
contains no persistent site, which accounts for the
strict inequality $\psi(r) < \overline{\theta}(r)$ found here
for $r>0$.
\label{figbound} }
\end{figure}

In the end we note that one can generalize the bound:
\begin{eqnarray}
\psi_{\hbox{pure}} \le \theta_{\hbox{pure}}
\end{eqnarray}
discussed in \cite{krapivsky_benaim}, to the disordered
case, as:
\begin{eqnarray}
\psi(r) \le \overline{\theta}(r)
\end{eqnarray}
i.e that 
the exponent $\psi(r)$ is always bounded by the persistence exponent 
$\overline{\theta}(r)$ of thermal
averaged trajectories found in (\ref{eqthetabarre}). This comes from
the observation that a point that has never been crossed by any particle
up to time $t$ for the effective dynamics
has to belong to domain that has a descendant still living at time t.
Here the reverse inequality is clearly not true (for
$p_0^* \neq 0$) since a surviving domain may not contain any
persistent site, as it can shift from its initial position,
as shown in figure \ref{figbound}. In particular we have found 
(\ref{eqthetabarre2}):
\begin{eqnarray}
\overline{\theta}(r)=1- 2 r +o(r)
\end{eqnarray}
Thus $\psi(r)$ and $\overline{\theta}(r)$ differ already 
at first order in $r$. This is different from the case of
the random field Ising model, studied in 
\cite{us_prl,us_long_rf}, where it is found that
$\psi=\overline{\theta}=(3-\sqrt{5})/4$, as 
the situation depicted in figure \ref{figbound}
does not occur.

\bigskip

\subsection{Statistics of coalescing particles}
\label{coales}

\begin{figure}[thb]
\centerline{\fig{9cm}{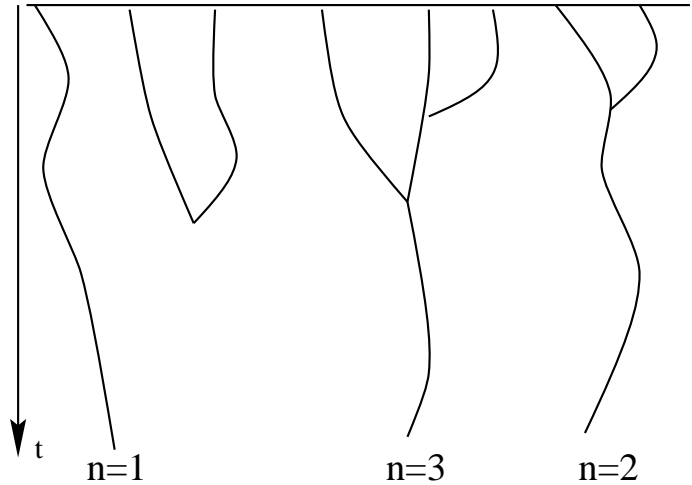}}
\caption{\narrowtext Number of ancestors of surviving
particle.
\label{figpart} }
\end{figure}

We now come to the study of persistence properties
associated to a particle. 
Following the general framework presented in the previous section
for the study of domain merging statistics,
we now introduce the number $D_n(t)$ of particles $A$ at time $t$
which have $n$ particles A for ancestors in the initial condition.
This is illustrated on the figure (\ref{figpart}). This will lead us
to introduce two new exponents, $\delta_A$ and $\psi_A$,
which have not been computed (nor defined) in the pure case
and which we will compute here in the disordered model.

The total number $N(t)$ of particles at time $t$
is simply given by the sum $N(t)=\sum_n D_n(t) \sim (\overline{l}(t))^{-1}$.
The fraction of initial particles which have 
a descendant still alive at $t$
is given by $S_A(t) = \sum_n n D_n(t) = <n> N(t)$. 
Again, the decay of $D_1(t)$, i.e of the number of particles
which have encountered no other particles, and of
$S_A(t)$ define two new independant exponents
\begin{eqnarray}
&& D_1(t) \sim \overline{l}(t)^{- \delta_A} \\
&& S_A(t) \sim \overline{l}(t)^{- \psi_A}
\end{eqnarray}
Together with these exponents, it is expected that
$D_n(t)$ takes the scaling form 
\begin{eqnarray}
D_n(t) = \frac{1}{\overline{l}(t)^{2-\psi_A} }{\cal D}
\left(\frac{n}{\overline{l}(t)^{1-\psi_A}}\right)
\end{eqnarray}
where the scaling function behaves for small $z$ as
${\cal D}(z) \sim z^{\sigma_A}$ for small $z$ where the exponent $\sigma_A$
is related to $(\delta_A,\psi_A)$ by the relation 
$\delta_A=2 - \psi_A+ (1-\psi_A) \sigma_A$.

We will compute these exponents via the RSRG by two methods.
The first one is direct, while the second one, presented at the
end, will rely on results previously established in 
Section \ref{asymptoticdynamics}.

In the first method, we introduce
an auxiliary variable $n$ for each valley containing
a particle A that counts the number of ancestors of this particle.
We then introduce the probability $P_A^{\Gamma}(\eta,\eta',n)$
that a valley at scale $\Gamma$ has $(\eta,\eta')$ and contains a particle A
having $n$ ancestors in the initial condition.
It satisfies the RG equation 
\begin{eqnarray}
&& \left[ \Gamma \partial_\Gamma - (1+ \eta) \partial_\eta 
- (1+ \eta') \partial_{\eta'} - 2 \right] P^{\Gamma}_A(\eta,\eta',n) \\
&& =   P^{\Gamma}_{A}(\eta,.,n) *_{\eta'} P^{\Gamma}_{0}(0, .) 
+ P^{\Gamma}_{A}(.,0,n) *_{\eta} P^{\Gamma}_{0}(., \eta') \\
&& +  P^{\Gamma}_{0}(\eta,.) *_{\eta'} P^{\Gamma}_{A}(0, .,n) 
+ P^{\Gamma}_{0}(.,0) *_{\eta} P^{\Gamma}_{A}(., \eta',n)  \\
&& + (1-r) \left[ P^{\Gamma}_{A}(\eta,.,.) *_{\eta',n} P^{\Gamma}_{A}(0, .,.) 
+ P^{\Gamma}_{A}(.,0,.) *_{\eta,n} P^{\Gamma}_{A}(., \eta',.)  \right]
\end{eqnarray}
where $P_0^{\Gamma}(\eta,\eta')$ is the probability that a valley at scale $\Gamma$ has $(\eta,\eta')$ and contains no particle, i.e. it satisfies
(\ref{rgreadiff}). At this stage, the variable $n$ is an integer
$n=1,2,3, \cdots$, and the convolution on $n$ is a discrete convolution.

At large $\Gamma$, we know the fixed point $P_0^*(\eta,\eta')=p_0^*
e^{-\eta-\eta'}$ and $\sum_{n} P_A^{\Gamma}(\eta,\eta',n)=p_A^*
e^{-\eta-\eta'}$ where $p_0^*=\frac{r}{r+1}=1-p_A^*$.
Thus setting $P_A^{\Gamma}(\eta,\eta',n)=p_A^* e^{-\eta-\eta'} \Delta^{\Gamma}(\eta+\eta',n)$,
we find that the function $\Delta^{\Gamma}(z,n)$ satisfies
\begin{eqnarray}
 \left[ \Gamma \partial_\Gamma - (2+ z) \partial_z  +z  \right] \Delta^{\Gamma}(z,n)  
=  2 p_0^* \int_0^z dz'\Delta^{\Gamma} (z',n) 
+(1-r) p_A^*  \Delta^{\Gamma} (.,.) *_{z,n} \Delta^{\Gamma} (.,.)
\end{eqnarray}
and we now will compute successively $\delta_A(r)$, $\psi_A(r)$ for all $r$
and the scaling function for $r=0$.

\subsubsection{  Exponent $\delta_A(r)$ for the process (\ref{potts})}

To compute the exponent $\delta_A$ giving the decay of the number
of particles A that have only one ancestor in the initial
condition, we need to solve the previous
equation for the particular value $n=1$.
$\Delta^{\Gamma}(z,n=1)$ decouples from the other values of $n$,
and satisfies the linear equation
\begin{eqnarray}
 \left[ \Gamma \partial_\Gamma - (2+ z) \partial_z  +z  \right] \Delta^{\Gamma}(z,1)  
=  2 p_0^* \int_0^z dz'\Delta^{\Gamma} (z',1) 
\end{eqnarray}
Since $D_{n=1}(t) = N(t) \int_{\eta,\eta'} P_A^{\Gamma} (\eta,\eta',1)
\sim \frac{1}{\Gamma^2} \int dz  z e^{-z} \Delta^{\Gamma}(z,1) $,
we set $\Delta^{\Gamma}(z,1)=\Gamma^{2-2 \delta_A} \Delta(z)$.
The function $\Delta(z)$ is then the solution of the differential equation
\begin{eqnarray}
(2+z) \Delta''(z) + (2 \delta_A -1 -z) \Delta'(z)-(1-2 p_0^*) \Delta (z)=0
\end{eqnarray}
together with the boundary condition at $z=0$
\begin{eqnarray}
2 \Delta'(0) + (2 \delta_A -2 ) \Delta(0)=0
\end{eqnarray}
We thus find $\Delta (z) \propto U(1-2 p_0^*,1+2 \delta_A,2+z)$, and the exponent
$\delta$ is determined by the implicit equation
\begin{eqnarray}
2  U'(1-2 p_0^*,1+2 \delta_A,2) +(2 \delta_A-2) U(1-2 p_0^*,1+2 \delta_A,2)=0
\end{eqnarray}
Using the functional relation $z U'(A,B,z)+(B-1-z) U(A,B,z)=-U(A-1,B-1,z) $,
and $p_0^*=\frac{r}{r+1}$, the exponent $\delta_A(r)$ is finally
the solution of the equation
\begin{eqnarray}
 U(- \frac{ 2 r}{r+1},2 \delta_A(r),2) =0
\label{eqdeltaa}
\end{eqnarray}
For the particular case $r=1$ where particles always annihilate upon meeting,
we have $\delta_A(r=1)=1$ as it should since in this case the particles can have 
only one ancestor $D_n(t)=\delta_{n,1} N_A(t)$.
In the limit $r \to 0$, where particles always coalesce upon meeting, we have
$\delta_A(r=0) \to + \infty$ : indeed in this case at large $\Gamma$,
all valleys contain a particle A $(p_A^*=1)$, 
and the probability to have $n=1$ decays exponentially with $\Gamma$,
since it requires that four consecutive bonds ( the two bonds of the
valley and the two neighbors ) are not decimated.

\subsubsection{  Exponent $\psi_A(r)$ for the process (\ref{potts})}

To compute the exponent $\psi_A$, we introduce the rescaled variable 
$\nu=\frac{n}{\Gamma^{2(1-\psi_A)}}$. The fixed point solution
$\Delta(z,\nu)=\frac{P_A^*(z,\nu)}{p_A^* e^{-z}}$ 
of the rescaled variables has to satisfy
\begin{eqnarray} \label{eqDelta}
&&  \left[ (2+ z) \partial_z  -z +2 (1-\psi_A) (\nu \partial_{\nu} +1) \right] \Delta (z,\nu) \\
&&  +  2 p_0^* \int_0^z dz'\Delta (z', \nu) 
+(1-r) p_A^*  \Delta (.,.) *_{z,\nu} \Delta (.,.) =0
\end{eqnarray}
In particular, using $p_0^*=\frac{r}{r+1}=r p_A^*$, we find that its first moment 
$C(z)= \int_0^{\infty} d\nu \nu \Delta(z,\nu)$
satisfies the differential equation 
\begin{eqnarray}
(2+ z) C''(z) +(2 \psi_A-1-z) C'(z) +(1-2p_0^*) C(z)=0
\end{eqnarray}
with the boundary condition at $z=0$
\begin{eqnarray}
2 C'(0) + 2 (\psi_A-1) C'(0))=0
\end{eqnarray}
So finally $C(z) \propto U(-1+2 p_0^*,1+2 \psi_A,2+z)$, where the exponent $\psi_A$
is the solution of the implicit equation
\begin{eqnarray}
2  U'(-1+2 p_0^*,1+2 \psi_A,2) + 2 (\psi_A-1) U(-1+2 p_0^*,1 + 2 \psi_A,2)=0
\end{eqnarray}
Using again the functional relation 
$z U'(A,B,z)+(B-1-z) U(A,B,z)=-U(A-1,B-1,z) $,
and $p_0^*=\frac{r}{r+1}$, the exponent $\psi_A(r)$ is finally
the smaller solution of the equation
\begin{eqnarray}
 U(-  \frac{2}{1+r}, 2 \psi_A(r),2) =0
\label{eqpsia}
\end{eqnarray}

For the particular case $r=1$ where particles always annihilate upon meeting,
we have $\psi_A(r=1)=1=\delta(r=1)$ as it should,
 since in this case the particles can have only one ancestor.
In the case $r = 0$, where particles always coalesce upon meeting, we have
$\psi_A(r=0) =0$ : indeed the probability for an initial particle
to have a descendant living at $\Gamma$ is one, and thus $S_A(t)$
is constant and not decaying.

\subsubsection{ Scaling function }

The distribution ${\cal D}(\nu)$
of the rescaled variable $\nu=n/\Gamma^{(2-2\psi_A)}$ can in principle be obtained 
in terms of the solution $\Delta(z,\nu)$ of (\ref{eqDelta}) as
\begin{eqnarray}
{\cal D}(\nu)=\int_0^{\infty} d\eta_1 \int_0^{\infty} d\eta_2 
\frac{P_A^*(\eta_1,\eta_2,\nu)}{p_A^*}
= \int_0^{\infty} dz z e^{-z} \Delta(z,\nu) 
\end{eqnarray}
In Laplace with respect to $\nu$, we have that $\hat{\Delta}_r(z,q)
=\int_0^{\infty} d\nu e^{-q \nu} \Delta_r(z,\nu)$ satisfies
\begin{eqnarray} \label{eqscalingfunction}
&&  \left[ (2+ z) \partial_z  -z -2 (1-\psi_A(r)) q \partial_q \right] 
\hat{\Delta}_r(z,q)  \\
&& +  \frac{2r}{1+r} \int_0^z dz' \hat{\Delta}(z',q) 
+\frac{1-r}{1+r}  \hat{\Delta}_r(.,q) *_{z} \hat{\Delta}_r(.,q) =0
\end{eqnarray}

In the case $r = 0$, where particles always coalesce upon meeting,
the number $n$ of ancestors should have the same statistical
properties as the length of a valley, and thus using the fixed
point solution (\ref{solu,fixedpointeta2}), we should have 
\begin{eqnarray} 
\int_0^{\infty} d\nu e^{-q \nu} \frac{P_A^*(z,\nu)}{p_a^*}
= \frac{q}{\sinh^2 \sqrt{q} } e^{-z \sqrt{q} \coth \sqrt{q} }
\end{eqnarray}
Indeed we find:
\begin{eqnarray} 
\hat{\Delta}_{r=0}(z,q)= \frac{q}{\sinh^2 \sqrt{q} } e^{z (1- \sqrt{q} \coth \sqrt{q} )}
\end{eqnarray}
is the solution of (\ref{eqscalingfunction}) 
for $(r=0,\psi_A=0)$, and thus in this case the scaling
function ${\cal D}(\nu)$ reads
\begin{eqnarray}
&& {\cal D}_{r=0}(\nu)= LT_{q \to \nu}^{-1} \left( \frac{1}{\cosh^2 \sqrt{q}}
\right) = \sum_{j=-\infty}^{+\infty}
(2 \nu \pi^2 (j+1/2)^2 - 1) 
e^{- \nu \pi^2 (1/2+ j )^2}  \\
&& = \frac{2}{ \sqrt{\pi} \nu^{3/2} }
\sum_{k=-\infty}^{+\infty} (-1)^{k+1} k^2  e^{- \frac{k^2}{\nu} } 
\end{eqnarray}

\subsubsection{second method to compute $\delta_A(r)$ and $\psi_A(r)$}

To compute $D_1(t)$, i.e the probability that a given particle
$A$ has met no other particles up to time $t$, we can consider 
this particle $A$ as a tagged particle, give it a name, say X,
and consider it as a new specie in very small concentration.
It satifies the following reaction rule:
\begin{eqnarray}
A + X \to 0 \qquad \text{proba}=1
\end{eqnarray}
and of course $X+0 \to X$ and the same reactions for the $A$
as before. We need only to work to linear order in $p_X$
and we are back exactly in the case studied in the section 
\ref{asymptoticdynamics} 
of the dynamics near the asymptotic state, of a new reaction diffusion
(whose fixed point has $p^*_X=0$). The corresponding 
eigenvalue of the matrix $M$ introduced in 
(\ref{matrixM}) is $\mu=p^*_0=r/(1+r)$. Here $0 \le \mu \le 1/2$ 
which corresponds to an attractive fixed point at $p_X=0$ (since the $X$
disappear) and with $p_X \sim \Gamma^{- \Phi}$ where $\Phi$ is 
solution of (\ref{eqphi}) for $\mu=r/(1+r)$. Since $D_1(t) \sim 
p_X/\Gamma^2 \sim \Gamma^{- 2 \delta_A}$ we recover the equation
(\ref{eqdeltaa}).

To compute the exponent $\psi_A(r)$ we need to consider similarly the
reaction for the tagged particle $X$:

\begin{eqnarray}
&& A + X \to X \qquad \text{proba}=1-r \\
&& A + X \to 0 \qquad \text{proba}=r 
\end{eqnarray}
and of course $X+0 \to X$ and the same reactions for the $A$.
In this case $\mu=p^*_0 + p^*_A (1-r) = 1/(1+r)$ and 
$1/2<\mu<1$ which corresponds to an {\it unstable}
fixed point at $p_X=0$. One finds $p_X \sim \Gamma^{- \Phi}$ where $\Phi$ is 
solution $\Phi_{-}$ of (\ref{eqphi}) for $\mu=1/(1+r)$. Since $S_A(t) \sim 
p_X/\Gamma^2 \sim \Gamma^{- 2 \psi_A}$ we recover the equation
(\ref{eqpsia}) which determines $\psi_A(r)$.

\section{disorder in the reaction probabilities}

It is interesting to study the stability of our results to
an additional quenched disorder in the reaction probabilities 
given by the matrix $W$ (i.e spatial inhomogeneities).
We continue to consider only the rule
that species react immediately when they encounter, but the analysis in fact
also covers - in an effective way - the case where reaction rates
are finite and with quenched disorder.
We sketch in this Section a possible way of
applying the present RSRG procedure to this case. 

Let us consider a model where the reaction probabilities
are themselves functions of the position 
$W^k_{k_1,k_2}(x)$. A simple example is to allow
the parameter $r$ to depend on $x$ as $r(x)$ in the
process (\ref{potts}). Let us examine what happens
at a decimation time scale $\Gamma=T \ln t$. The particle in state
$k_1$ in the decimated valley jumps over the barrier
to a valley containing $k_2$. Since $k_2$ is typically
at equilibrium at the bottom of its valley, the reaction 
is most likely to take place at the bottom of the valley 
within a $O(1)$ distance of it
(since this is where all the weight of the particule 
$k_2$ is concentrated). Thus as time increases, the total number
of sites in the system where reactions can typically occurs decays as $1/\Gamma^2$.
In each renormalized valley at $\Gamma$ there is typically a ``finite''
number (i.e not growing with $\Gamma$) of sites $x$ where reactions
occur and thus a ``finite'' number of possible values $W(x)$
(a notation for the set of $W^k_{k_1,k_2}$). For each
valley these form a given set fixed in time. There are thus a priori
two competing effects: the several values taken by $W$ in a valley result in
an ``averaging'' effect for the effective $W$ of this valley.
On the other hand the fact that this set is finite and fixed in
time implies non trivial correlations between two reactions
occuring at different times in the same valley.

Here we will restrict to consider a toy model
where we assign a single transition matrix $W$ to each renormalized 
valley with probability $P(W)$. It would be accurate in the case 
where in the initial distribution the $W$'s are correlated over distances 
much larger than the typical thermal width of a packet $\sim O(1)$, but still 
small compared to $\Gamma^2$. 
This problem can thus be treated
by introducing the probability $P_k(\eta_1,\eta_2,W)$
that a valley has rescaled barriers $\eta_1$, $\eta_2$ and
an associated rate $W$. When two valley merge upon a decimation
the new one simply keeps the $W$ of the lowest one.
One notes that the statistical independence
is again preserved by RG. The RG equation is simply
\begin{eqnarray}
&& \Gamma \partial_\Gamma P^{\Gamma}_k(\eta_1,\eta_2,W)
=((1+\eta_1) \partial_{\eta_1} + (1+\eta_2) \partial_{\eta_2} ) 
+ 2) P^{\Gamma}_k(\eta_1,\eta_2,W) \nonumber \\
&& + W^k_{k_1,k_2}
[ P^{\Gamma}_{k_1}(\eta_1,.,W) *_{\eta_2} P^{\Gamma}_{k_2}(0, .) 
+ P^{\Gamma}_{k_1}(.,0) *_{\eta_1} P^{\Gamma}_{k_2}(., \eta_2,W) ]
\label{rgdisorder}
\end{eqnarray}
where $\int_W P_k(\eta_1,\eta_2,W)=P_k(\eta_1,\eta_2)$
and summation over repeated indices is implied.
We also note that the distribution of $W$, $P_\Gamma(W)=\sum_k 
\int_{\eta_1,\eta_2} P_k(\eta_1,\eta_2,W)$ is preserved
by the RG rule, thus $P^\Gamma(W)=P(W)$. Thus we have a
``marginal'' problem, since in this toy model $P(W)$ 
does not flow by RG \cite{footmarginal}.

One can now look for fixed points of this RG equation
under the form:
\begin{eqnarray}
P_k(\eta_1,\eta_2,W) = e^{-\eta_1 - \eta_2} P_k(W) 
\end{eqnarray}
where the $P_k(W)$ must satisfy:
\begin{eqnarray}
P_k(W)  = W^k_{k_1,k_2} P_{k_2}(W) \int dW' P_{k_1}(W')
\end{eqnarray}

In the case of the model (\ref{potts}) with a distribution
$P(r)$ of $r$, one can show that a solution is:
\begin{eqnarray}
P_k(r)  = P(r) p_k^*(r) 
\end{eqnarray}
where $p_0^*(r)=r/(1+r)$ and $p_A^*(r)=1/(1+r)$ are the
equilibrium occupation probabilities for the problem with
a uniform $r$, solution of the equation 
$p^*_k(W)  = W^k_{k_1,k_2} p^*_{k_2}(W)  p^*_{k_1}(W)$. 
Such a simple solution holds in that case because of the
form of the matrix $M$ (\ref{matrixM}) which is simply a projector onto
the vector $p_k^*$. In general this does not hold and one has to 
solve the above equation. It is then possible in principle
to perform, for an arbitrary $P(W)$,
the same study as the one done here, such as stability eigenvalues around the
fixed point etc.., which is left for the future.

To summarize, the above result indicates that within the toy model and the effective
dynamics, quenched disorder in the transition matrix $W(x)$ will lead to a modification
of the large time properties. These properties can be computed using the RG by assigning a 
an effective reaction probability matrix of each valley. They depend in a 
continuous way on the asymptotic distribution $P(W)$. There is thus an
infinite dimensional manifold of fixed points in the RG sense, and the problem
is {\it marginal}. Like all marginal problems, it is very sensitive to
corrections which may make disorder marginally irrelevant or marginally relevant
(or remain strictly marginal). The averaging effect may in the end make 
the disorder marginally irrelevant, but to decide, within
the real dynamics \cite{footmarginal} and within a model with shorter disorder 
correlation length, whether disorder 
is actually marginally irrelevant, and how it flows,
requires a more detailed study which goes beyond this work.

\section{conclusion}

\label{conclusion}

In this paper we have studied the problem of various species
of particles diffusing in
presence of quenched random local bias (Sinai
landscape) and reacting upon meeting.
We have shown that the real space renormalization 
group method (RSRG), which has proved to be a
powerful tool to study single particle diffusion 
in Sinai landscape \cite{us_prl}, can be extended in a simple and
natural way to study a large class of reaction diffusion models.
Since here also the physics is controlled by infinitely broad disorder 
fixed points, this method, as in the single particle problem,
is expected to yield the {\it exact} large time behaviour.
Focusing on renormalized valleys as well as on the particles
(and species) contained in these valleys, and following the 
evolution of their distribution by decimation upon increase
of the time scale, allowed us to obtain many new exact 
results on this problem. 

We have obtained a detailed description of the asymptotic
states, such as the large time decay of the density of
each specie, $n_k(t)$, and the spatial distribution of particles.
It confirms that in $d=1$ Sinai landscape the reaction is
subdiffusion limited. The first step was to identify simple fixed points 
of the valley distribution RG equation,
which correspond - for a given 
reaction process described by a transition matrix - 
to possible asymptotic dynamical states. Each of these
states is characterized by fixed fractions $p^*_k$ for each specie,
the physical picture being the following. 
At time scale $\Gamma = T \ln t$ the system consists of a
set of successive renormalized valleys, which can be either
empty, with probability $p^*_{\emptyset}$ or contain a
particle of specie $k$, with probability $p^*_k$.
The separation between particles grows as the characteristic
length $\overline{l}(t) \sim \Gamma^2$, and thus
$n_k(t) \sim p_k^*/(T \ln t)^2$. The decay of concentration,
when compared to the single particle diffusion length, leads
to define universal amplitudes, which we obtained exactly.
Also, from the exact statistical
independence of the successive valley lenghts, the distribution
of interval between particles (domains) was derived
(and compared with some pure case results).

To confirm that a given fixed point is indeed an asymptotic state,
actually reached by the system at large time,
it is necessary to study its linear stability.
We have thus obtained analytically the spectrum of stability eigenvalues 
around any simple fixed point, as a function of the reaction transition matrix, 
thereby solving the stability problem. The convergence towards these asymptotic
states (i.e the attractive RG fixed points) was studied.
The leading convergence towards these asymptotic states was found
to be generically as $|p_k(t)-p_k^*| \sim (T \ln t)^{-\Phi}$
with a non trivial $\Phi$, solution of a hypergeometric equation
(with in addition, an amplitude periodic in $\ln \ln t$
in the case of complex eigenvalues). In some cases, 
corresponding to $\Phi = +\infty$, the convergence is faster
as a power law in time with non universal exponent
depending on details of the initial model.

Eigenvalues corresponding to unstable fixed points,
which were also determined, are of particular interest 
for reactions which lead to several distinct dynamical phases
(i.e several possible asymptotic states). The
transitions between different dynamical phases
being controlled by such unstable RG fixed points,
we have thus obtained the corresponding critical
exponents. As an example, a process with a non trivial
phase diagram was studied.

We have also studied persistence properties associated to
a given asymptotic state. As in pure systems, where it was
originally defined, persistence can be studied for various types
of patterns (single particles, domains etc..). Remarkably,
for the disordered models at hand we are able to derive a {\it much larger}
set of exact results than exists at present for the corresponding 
pure systems. 

We have first obtained, for a generic process,
the decay exponent $\theta$ for the probability of
no crossing of a given point by single particle trajectories. As
noted in \cite{us_prl}, in a disordered system, persistence
of thermal averages can be quite different from 
single particle persistence. Thus we have also 
computed the probability of no crossing of a given point by 
thermally averaged packets,
which yields the decay exponent $\overline{\theta}$.
The properly generalized persistence exponents
associated to $n$ crossings have been defined, and computed.
Next, we have characterized the statistics of domains,
which can disappear or merge as time increases.
Restricting, for simplicity, to
the process $A+A \to \emptyset$ or $A$ with probabilities
$(r,1-r)$, we have obtained exactly the exponents $\delta(r)$
and $\psi(r)$ characterizing the survival up to time $t$
of a domain without any merging or with mergings respectively.
We have also introduced new exponents which similarly characterize
the statistics of the coalescence of particles. We have then
computed them, namely $\delta_A(r)$ and $\psi_A(r)$ 
characterizing the survival up to time $t$
of a particle $A$ without any coalescence or with coalescences 
respectively.

We have found these new exponents as solutions of  
hypergeometric equations. For comparison,
the only known analytical result
in the pure case is for the exponent $\theta_{pure}(r)$
for the process (\ref{potts}). A surprising outcome
was that several exact exponents of the model with disorder
were found to be numerically very close, for all values of $r$, 
to some exponents for the pure system, although
they are associated to completely
different diffusion length ($l_{pure} \sim \sqrt{t}$ while
$\overline{l}(t) \sim (T \ln t)^2$). Indeed we found that 
$\overline{\theta}(r) \approx \frac{1}{2} \theta_{pure}(r)$
although they are definitely distinct, and furthermore that
$\psi(r) \approx \psi_{pure}(r)$ and 
$\delta(r) \approx \delta_{pure}(r)$ where 
$\psi_{pure}(r)$, $\delta_{pure}(r)$ - not known analytically -
are extracted from the numerical simulation of \cite{krapivsky_benaim}.
The agreement in relative values is better than about one percent
for all $r$. 
It may be that this observed numerical coincidence
could be traced to the exact statistical independence of valley 
lengths in the disordered problem, while 
the so called ``independent interval approximation'',
gives reasonable approximation in the pure case (but, surprisingly, poorer
than the one provided by these new exponents).
This however is far from an explanation and further investigation
may be of interest. 

The effect of additional disorder in 
the reaction rates was also discussed. In a simple case 
it was found to be marginal, and thus 
yield non trivial modifications, continuously varying with the
disorder distribution. The question of whether this
result is stable to corrections resulting from the real dynamics or
from disorder with shorter correlation length remains to be further investigated

Although we have not considered explicitly branching BARW processes, with
additional creation of particles, it is clear that for at least some
of them the physics should not be qualitatively too different from
the one obtained here. Indeed, since in Sinai disorder particles
are essentially confined to the bottom of large renormalized wells,
as long as the process is such that particles are not created out of the
vacuum and that the annihilation reactions are sufficient to maintain
the number of particles small when at local equilibrium in a well,
the reaction can be treated very similarly via RSRG as for the
model studied here. We have thus characterized a broad set of
reaction diffusion models with disorder.

Finally, it is worthwile to mention
that we have also identified cases which clearly require
a more complicated analysis going beyond the
present paper. For instance we have given
an example of a marginal reaction, which requires a
non linear stability analysis. Also, we have given 
an example of a cyclic reaction for which all simple RG fixed points 
are shown to be unstable. The question of the determination of the true 
asymptotic states of this reaction is thus still open. Another open 
interesting, and maybe related, question is
whether reactions with large enough number of species, which can
lead to chaotic attractors in pure cases
\cite{velikanov_chaotic,nicolis_chaos},
will also lead to chaotic behaviour in presence of
disorder.

\begin{figure}%
\tabcolsep2.pt
\renewcommand{\arraystretch}{3.}
\noindent
\begin{tabular}{c||c|c}
 $\text{Event}$ & $\text{Exponent}$ & $\text{Equation}$ \\ 
\hline \hline 
$\text{no crossing of O by any particle}$ & $\theta(r)$  & $\theta = 1/(1+r)$ 
\\  \hline
$\text{$n=g \ln \Gamma$ particles absorbed at O}$ 
& $\omega(g)$  & $2 \omega = (1+r)^{-1} - g 
+ g \ln((1+r)g)$ 
\\  \hline
$\text{no crossing of O by thermal. aver. traj.}$ & $\overline{\theta}(r)$  
& $\overline{\theta} 
U(- \frac{r}{1+r}, 2 \overline{\theta},1) 
= U(- \frac{r}{1+r}, 2 \overline{\theta} + 1,1)$ 
\\  \hline
$\text{domain survival without merging}$ & $\delta(r)$  & 
$U(- 1/(1+r), 1+ \delta,2) =0$ 
\\  \hline
$\text{domain survival with merging}$ & $\psi(r)$  & 
$U(-\frac{1}{2}-\frac{r}{1+r} -\frac{\sqrt{1+6r+r^2}}{2 (r+1)},2 \psi,2)= 0$ 
\\  \hline
$\text{particle survival without coalescence}$ & $\delta_A(r)$  & 
$U(- 2 r/(1+r) ,2 \delta_A,2) =0$ 
\\  \hline
$\text{particle survival with coalescence}$ & $\psi_A(r)$  & 
$U(-  2/(1+r), 2 \psi_A ,2) =0$ 
\\ 
\hline
\end{tabular}

\vspace{5mm}\noindent
\small {\bf Table I}:
{Summary of the results obtained in the text for persistence exponents associated to
the decay of the probability of each indicated event, in the case of the
reaction diffusion process $A + A \to \emptyset$ (proba $r$) $A + A \to A$ (proba $1-r$).}
\end{figure}

\bigskip

{\bf Acknowledgements}

We thank D. S. Fisher for fruitful discussions
as well as S. Fauve, U. Tauber and P. Chauve for helpful 
remarks.

\appendix

\section{ Auxiliary variable for valleys }

\label{auxiliary}

In this Appendix we study auxiliary variables $(m)$
associated to bonds that evolve upon decimation as follows.
Consider the decimation of bond $(2)$ on Fig. \ref{fig2}: the two valleys 
corresponding to bonds $(1,2)$ and $(3,4)$ and containing 
respectively the species $k_1$ and $k_2$ merge, and 
the specie $k_1$ jumps to the bottom of valley $(3,4)$
and thus goes over the bond $(2)$ and $(3)$ to react there with the
specie $k_2$. It is thus natural to consider an auxiliary variable
$m$ which, upon decimation of the barrier $F_3'=F_1+F_3-\Gamma$,
evolves with the general rule:

\begin{eqnarray}  \label{rulempersistence}
m_3' = d_{k_1} m_1 + b_{k_1} m_2 + a_{k_1} m_3
\end{eqnarray}
where the coefficients $(a_k,b_k,d_k)$ depend on the specie $k$
which diffuses upon the corresponding decimation.

We now write the valley RG equation for 
$P^{\Gamma}_k(\eta,\eta',m,m')$ :

\begin{eqnarray}
&& (\Gamma \partial_\Gamma - (1+ \eta) \partial_\eta 
- (1+ \eta') \partial_{\eta'} - 2)
P^{\Gamma}_k(\eta,\eta',m,m') \\
&& = 
 W^k_{k_1,k_2} \int_{m_1,m_1',m_2,m_2'} \\
&& [  P^{\Gamma}_{k_1}(\eta,.,m,m_1') *_{\eta'}
 P^{\Gamma}_{k_2}(0, .,m_2,m_2') 
  \delta(m' - a_{k_2} m_1' - b_{k_2} m_2 - d_{k_2} m_2')  \\
&& + P^{\Gamma}_{k_1}(.,0,m_1,m_1') *_{\eta} 
P^{\Gamma}_{k_2}(., \eta',m_2,m') 
\delta(m - d_{k_1} m_1 - b_{k_1} m_1' - a_{k_1} m_2)]
\end{eqnarray}
Integrating this equation over the $m$ variables one
recovers of course the specie valley RG equation (\ref{rgreadiff}).

We now define the first moment :

\begin{eqnarray}
&& G_k^{1}(\eta_1,\eta_2) = \int_{m_1,m_2} m_1 P_k(\eta_1,\eta_2,m_1,m_2)  \\
&& G_k^{2}(\eta_1,\eta_2) = \int_{m_1,m_2} m_2 P_k(\eta_1,\eta_2,m_1,m_2)
\end{eqnarray}

Since we are looking at the symmetric case, we have
that $G_k^{2}(\eta_1,\eta_2) = G_k^{1}(\eta_2,\eta_1)$. We can thus
write the following closed equation for $G^{1}_k(\eta_1,\eta_2)$

\begin{eqnarray}
&& (\Gamma \partial_\Gamma - (1+ \eta_1) \partial_{\eta_1}
- (1+ \eta_2) \partial_{\eta_2} - 2) G^{1}_k(\eta_1,\eta_2) \\
&& =
W^k_{k_1,k_2} [ G^{1}_{k_1}(\eta_1,.) *_{\eta_2} P_{k_2}(0, .) 
+ 
a_{k_1} P_{k_1}(.,0) *_{\eta_1} G^{1}_{k_2}(., \eta_2) \\
&& +
b_{k_1} G^{1}_{k_1}(0,.) *_{\eta_1} P_{k_2}(.,\eta_2)
+
d_{k_1} G^{1}_{k_1}(.,0) *_{\eta_1} P_{k_2}(.,\eta_2)
\end{eqnarray}

In the asymptotic state we use the fixed point solution
$P^*_k(\eta_1,\eta_2)=p_k^* e^{-\eta_1-\eta_2}$, and write 
$G^{1}_k(\eta_1,\eta_2) = e^{-\eta_1-\eta_2}
g_k(\eta_1,\eta_2)$ and obtain the equation for the new function $g_k$:

\begin{eqnarray}
&& \Gamma \partial_\Gamma g_k(\eta_1,\eta_2) =
[ (1+ \eta_1) \partial_{\eta_1} - \eta_1 + (1+ \eta_2) \partial_{\eta_2} - \eta_2] g_k (\eta_1,\eta_2) \\
&& + 
W^k_{k_1,k_2} [ p_{k_2}^* \int_0^{\eta_2} g_{k_1}(\eta_1,.)
+  p_{k_1}^* a_{k_1} \int_0^{\eta_1} g_{k_2}(.,\eta_2) \\
&& + p_{k_2}^* b_{k_1} \int_0^{\eta_1} g_{k_1}(0,.)
+ p_{k_2}^* d_{k_1} \int_0^{\eta_1} g_{k_1}(.,0)
\end{eqnarray}

Since the $m$ variable is associated to bonds,
it is natural to look for solutions
where the function $g_k(\eta_1,\eta_2)$
is a function of $\eta_1$ alone.
We thus try solutions of the form
$g^{\Gamma}_k(\eta_1,\eta_2) = \Gamma^{\psi} g_k(\eta_1)$,
where the exponent $\psi$ characterizes the scaling of the
$m$ variable $m \sim \Gamma^{\psi}$.

For this to work we obtain, in terms of the matrix $M$ defined in (\ref{matrixM}),
the necessary condition:

\begin{eqnarray} \label{condition1}
g_k (\eta) = W^k_{k_1 k_2} p^*_{k_2} g_{k_1} (\eta)=M_{k,k_1} g_{k_1} (\eta)
\end{eqnarray}
together with the differential equation for $g_k(\eta)$:

\begin{eqnarray}
&& 0 = [(1 + \eta) \partial_{\eta} - \eta - \psi] g_k(\eta) 
+(M_{k,k_1} d_{k_1} + W^k_{k_1,k_2} p^*_{k_2} a_{k_2})
\int_0^{\eta} g_{k_1}(.) + M_{k,k_1} b_{k_1} g_{k_1}(0) \eta 
\end{eqnarray}

One can then try $g_k(\eta) = p^*_k \psi(\eta)$, which
automatically satisfies the necessary condition (\ref{condition1})
above (since $p_k^*$ is by construction an eigenvector of the $M$ matrix
of eigenvalue 1), and then the second equation gives
the conditions involving two numbers $\lambda_{1,2}$

\begin{eqnarray}  \label{condition2}
&&M_{k,k_1} p_{k_1}^* (d_{k_1}+a_{k_1}) = \lambda_1 p^*_k \\
&& M_{k,k_1} p_{k_1}^* b_{k_1}= \lambda_2 p^*_k 
\end{eqnarray}
together with the differential equation for $g(\eta)$:
\begin{eqnarray} \label{eqgpersist}
&& 0 = [(1 + \eta) \partial_{\eta} - \eta - \psi] g(\eta) 
+ \lambda_1 \int_0^{\eta} \psi(.) + \lambda_2 \psi(0) \eta
\end{eqnarray}

We now give two applications of this general analysis.

\subsection{Persistence exponent $\overline{\theta}$}

We now study the case $a_k=b_k=\delta_{k,0}$ and $d_k=1$ 
corresponding to the auxiliary variable (\ref{rulempersistence})
needed to compute the persistence exponent $\overline{\theta}$ 
The conditions (\ref{condition2}) become:
\begin{eqnarray}
&& p_k^*+M_{k,0} p_{0}^*  = \lambda_1 p^*_k \\
&& M_{k,0} p_{0}^* = \lambda_2 p^*_k 
\end{eqnarray}
Since the rates involving the empty state $(0)$
are given by definition by $W^k_{i 0} = \delta_{k,i}$,
we have $M_{k,0} =W^k_{0,i} p_i^* =\delta_{k,i} p_i^*=p_k^*$.
The conditions above are thus satisfied with 
$\lambda_1=1+p^*_0$ and $\lambda_2=p^*_0$ and thus the 
problem reduces to studying the 
the equation (\ref{eqgpersist}) for $g(\eta)$:

\begin{eqnarray}
 0 = [(1 + \eta) \partial_{\eta} - \eta - \psi] g(\eta) 
+ (1 + p^*_0) \int_0^{\eta_1} g(.) + p^*_0 g(0) \eta_1
\end{eqnarray}
Differentiating with respect to $\eta$ one gets:

\begin{eqnarray}
0 = (1+\eta) \partial^2_\eta g(\eta) + (1- \phi  - \eta) \partial_\eta g(\eta) 
+ p^*_0 (g(\eta) + g(0))  
\end{eqnarray}
with the boundary condition $g'(0)=\psi g(0)$. 
The solution of this confluent hypergeometric equation 
that is well behaved solution at infinity (i.e not
growing exponentially) reads:
\begin{eqnarray}
g(\eta)= g(0)
( 2 \frac{U(- p_0^*, 2 - \phi, 1 + \eta)}{U(- p_0^*, 2 - \phi, 1)} -1 )
\end{eqnarray}
The boundary condition at $\eta=0$ then leads to the following equation
for the exponent $\psi$ governing the scaling of the m variable 
$m \sim \Gamma^{\psi}$, as a function of $p_0^*$:
\begin{eqnarray}
U'(- p_0^* , 2 - \psi ,1) 
= \frac{\psi}{2} U(- p_0^* , 2-\psi,1)
\end{eqnarray}

Using functional relations of the confluent hypergeometric functions $U$,
we finally obtain that the fraction of sites that have never been
crossed by any particle in the effective dynamics decays as
$\frac{m}{\overline{l_{\Gamma}}} \sim \Gamma^{\psi-2} 
\sim (\overline{l_{\Gamma}})^{-\overline{\theta}}$
where the persistence exponent $\overline{\theta}=(2-\psi)/2$
is solution of the equation

\begin{eqnarray}
\overline{\theta}~~U(- p_0^* , 2 \overline{\theta},1) 
= U(- p_0^* , 2 \overline{\theta} + 1,1)
\end{eqnarray}

\subsection{Number of thermal packets seen by a given point}

We introduce the bond variable $m(n)$ which is the number of 
sites on the bond which have been crossed 
exactly $n$ times by a particle (any non empty state)
in the effective dynamics (i.e by a thermally averaged trajectory). 
It satifies upon decimation of bond $(2)$ with the same conventions 
as above:

\begin{eqnarray} 
&& m_3'(n) = m_1(n) + m_2(n-1) + m_3(n-1)  \qquad \hbox{if} \ k_1 \neq 0 \\
&& m_3'(n) = m_1(n) + m_2(n) + m_3(n)  \qquad \hbox{if}  \ k_1 = 0
\end{eqnarray}
Introduces the generating function
$m(z)=\sum_{n=0}^{n=+\infty} m(n) z^n$, the rule becomes: 

\begin{eqnarray} \label{rulemz}
m_3'(z) = d_{k_1}(z) m_1(z) + b_{k_1}(z) m_2(z) + a_{k_1}(z) m_3(z)
\end{eqnarray}
which, for fixed $z$, is the same rule as above with now
$a_{k}(z) = b_{k}(z)=\delta_{k,0} + z (1-\delta_{k,0})$
and $d_{k}= 1$.
The conditions (\ref{condition2}) become
\begin{eqnarray} 
&& (1+z) p_k^*+ (1-z) M_{k,0} p_{0}^*  = \lambda_1 p^*_k \\
&& z p_k^* +(1-z) M_{k,0} p_{0}^* = \lambda_2 p^*_k 
\end{eqnarray}
and thus using again $W^k_{i 0} = \delta_{k,i}$,
these conditions are satisfied with 
$\lambda_1(z)=1+z+(1-z)p^*_0$ and $\lambda_2=z+(1-z)p^*_0$.,
i.e. we only need to perform the substitution
$p^*_0 \to z + p^2_0 (1-z)$ in the previous solution
to obtain the equation for the exponent $\overline{{\cal \theta}}(z)$
governing the scaling of the ratio 
$\frac{m(z)}{\overline{l_{\Gamma}}} \sim (\overline{l_{\Gamma}})^{- 
\overline{{\cal \theta}}(z)}$ :

\begin{eqnarray} \label{thetaz}
\overline{{\cal \theta}}(z) ~~U(- z - p_0 (1-z) , 2
\overline{{\cal \theta}}(z),1) 
= U(- z - p_0 (1-z) , 1+2\overline{{\cal \theta}}(z) ,1)
\end{eqnarray}

The probability that a given point has
been visited by $n$ thermally averaged trajectories
up to time $t$, is thus obtained in 
the rescaled variable $g=\frac{n}{ \ln \Gamma}$
as:
\begin{eqnarray} 
Prob(g) \sim (\overline{l_{\Gamma}})^{- \overline{\theta}(g)}
\end{eqnarray}
It decays with the exponent $\overline{\theta}(g)$ obtained 
through the Legendre transform
\begin{eqnarray}
2 \overline{\theta}(g) = 2 \overline{{\cal \theta}}(z^*(g)) + g \ln (z^*(g))
\end{eqnarray}
where $z^*(g)$ is the solution of $2 \frac{\overline{{\cal \theta}}(z)}{dz}
+\frac{g}{z}=0$.

One can compute simply the value $g_a$ that $g$ takes with probability one as $\Gamma \to \infty$.
It is given by $\overline{\theta}(g_a)=0=\frac{\overline{\theta}(g)}{dg}\vert_{g=g_a}$.
This gives $g_a=-2 \frac{\overline{{\cal \theta}}(z)}{dz}\vert_{z=1}$,
and thus differentiating (\ref{thetaz}) with respect to $z$
and taking $z=1$ we finally get 
\begin{eqnarray}   
g_a= (1-p_0^*) \frac{U_1(-1,1,1)}{U(-1,0,1)/2-U_2(-1,1,1)} =  \frac{4}{3} (1-p_0^*)
\end{eqnarray}
where we have used the notations $U_1(a,b,z) \equiv \partial_a U(a,b,z)$ and 
$U_2(a,b,z) \equiv \partial_b U(a,b,z)$.

\section{ The particular case of ``associative processes" }

\label{appassoc}

It turns out to be useful to introduce the
notion of ``associative processes" :
these are processes such that the outcome of a sequence
of reactions does not depend on the order in which it was performed,
i.e. such that the rates $W$ satisfy:

\begin{eqnarray}
W^k_{p_1 \alpha} W^\alpha_{p_2 p_3} = W^k_{p_2 \beta} W^\beta_{p_1 p_3}
\label{associative}
\end{eqnarray}
for all $k,p_1,p_2,p_3$ (contraction over $\alpha$ and $\beta$ is implied).
This means that the probability of $(p_2.p_3).p_1 = k$ is identical
to the probability of $(p_1.p_3).p_2 = k$ ($a.b$ denotes the result of the
reaction of $a$ and $b$). For example, the process defined in (\ref{potts}) 
is associative.

An important property of associative processes is that their 
matrix $M$ \ref{matrixM} satisfies:

\begin{eqnarray}
M^2 = M
\end{eqnarray}
and thus the eigenvalues $\mu_\alpha$ have only two possible values : $0$ or $1$.

For the RG, these processes have also the following interesting property :
the subspace of valley distributions of the form

\begin{eqnarray}
P^{\Gamma}_k(\eta_1,\eta_2) = W^k_{r_1,r_2} 
P^{\Gamma}_{r_1}(\eta_1) P^{\Gamma}_{r_2}(\eta_2)
\end{eqnarray}
is conserved by the RG (\ref{rgreadiff}), provided that the bond distribution
$P^{\Gamma}_{k}(\eta)$ satisfies the bond RG equation

\begin{eqnarray}  \label{rgbonds}
\Gamma \partial_\Gamma P^{\Gamma}_k(\eta) = ( (1+\eta)
\partial_\eta  + 1 ) P^{\Gamma}_k(\eta) +
W^k_{k',k_3} W^{k'}_{k_1,k_2} P^{\Gamma}_{k_2} (0)
 P^{\Gamma}_{k_1}(.)*_{\eta} P^{\Gamma}_{k_3}(.)
\end{eqnarray}

The bond RG equation (\ref{rgbonds}) can in fact be interpreted
to characterize the following modified reaction diffusion process,
that we call ``the bond-reaction diffusion
process" : one associates to the bottom of
each bond (i.e the point of lowest energy) a specie in one of the
possible ``states'' and define probability distributions
 $P_k(z)$ for the bonds. We consider two consecutive valleys
made with the bonds (1,2) and (3,4). Initially the bonds (1,2,3,4) respectively
contain the species $k_1,k_2,k_3,k_4$. Upon decimation of
bond (2), the bond diffusion process
is defined as follows in three steps
(i) First the two species $k_1,k_2$ on bonds (1,2) 
react to give another state $k'$ with the rates $W^{k'}_{k_1,k_2}$
(ii) the new specie $k'$ 
diffuses towards the bottom of the bond 3.
(iii) the species $(k',k_3)$ react at the bottom of bond 3
 to give a new species $k_3'$ with the rates $W^{k_3'}_{k',k_3}$).
For comparison, it is useful to recall the corresponding
real dynamics with valleys : initially, the valley $(1,2)$
contains some species $k'$, the valley $(3,4)$ some species $k''$.
Upon decimation of bond (2), the species $k'$ diffuses towards
the bottom of the valley $(3,4)$ and reacts there with $k''$
to give $k$ with probability $W^k_{k',k''}$. Thus,
in the end, the physical content (the specie) of the renormalized valley
for the bond-diffusion process is $k$ with probability $W^k_{k_3',k_4}
W^{k_3'}_{k',k_3}$, whereas in the original valley process,
the final result is $k$ with probability $W^k_{k',k''} W^{k''}_{k_3,k_4}$.
The two descriptions are thus equivalent in that sense only if the rates
satisfy the associativity condition (\ref{associative}).


\begin{thebibliography}{10}
\bibitem{footnote4}
see however \cite{schutz_oned} and discussion below.
\bibitem{footgamma}
We will denote everywhere the dependence of all quantities in the variable
  $\Gamma$ by a superscript, which should not be confused for a power.
\bibitem{footnotel}
Note that the definition of the symbol $\overline{l}_\Gamma$ is different from
  the one used in \cite{us_long} where it denoted the averaged length of a
  valley.
\bibitem{footnotecardy}
Keeping in mind the cautionary remarks in \cite{footnoterg}, we can still
  attempt to compare with the $d=2-\epsilon$ expansion formula (14) from
  \cite{richardson_cardy99}, which reads ${\cal A}= 0.106103 \frac{1}{\epsilon}
  + 0.0263539 + O(\epsilon)$ and leads to ${\cal A} \sim 0.13$ by setting
  $\epsilon=1$.
\bibitem{footnoteng}
Restoring the microscopic time unit $t_0$, one notes that the coefficient $c$
  of the $O(1/\Gamma)$ correction in $n_\Gamma$ can be set to $0$ by a
  redefinition of the microscopic time scale $t_0$ \cite{footnotet0}. This
  redefinition does not affect the formula for the $p_k(t)$ as it corresponds
  to the landscape eigenvector with $\Phi=+1$.
\bibitem{footnotend}
The case of a non diagonalisable block of degenerate eigenvalues also deserves
  a separate study, not performed here. It is likely to lead to additional
  logarithmic corrections ($\ln \Gamma$ and powers of it) to the large time
  behaviour.
\bibitem{footmarginal}
Within this toy model there will be corrections due to rare events which split
  the thermal packet between distants wells. For instance events (a) in Fig. 7
  of \cite{us_long} tend to average the disorder in the $W$. However they occur
  with probability at most of order $O(1/\Gamma)$ and should not affect the
  leading behaviour. A detailed study of these corrections goes beyond this
  paper.
\end{thebibliography}

\end{document}